\pdfoutput=1
\documentclass[a4paper,11pt]{article}

\usepackage{lineno}
\usepackage{jheppub}
\usepackage[utf8]{inputenc}
\usepackage{amsmath}
\usepackage{amsthm}
\usepackage{amssymb}
\usepackage{enumitem}   
\usepackage[english]{babel}
\usepackage{url}
\usepackage{mathtools}
\usepackage{bbold}
\usepackage{slashed}
\usepackage{multirow}
\usepackage{lipsum}
\usepackage{xcolor}
\usepackage{float}
\usepackage{revsymb}
\usepackage{hyperref}

\newcommand{\sm}{\hspace{-1pt}}
\newcommand{\bm}[1]{\boldsymbol{#1}}

\begin{document}


\preprint{ACFI-T23-05, MS-TP-23-38}
\title{Axion inflation in the strong-backreaction regime:\\decay of the Anber--Sorbo solution}

\author[a]{Richard~von~Eckardstein,}
\author[b,c]{Marco~Peloso,}
\author[a]{Kai~Schmitz,}
\author[a,d]{Oleksandr~Sobol,}
\author[e]{and Lorenzo~Sorbo}

\affiliation[a]{Institute for Theoretical Physics, University of M\"{u}nster, \\Wilhelm-Klemm-Stra{\ss}e 9, 48149 M\"{u}nster, Germany}
\affiliation[b]{Dipartimento di Fisica e Astronomia ``Galileo Galilei'', Universit\`{a} degli Studi di Padova,\\Via Marzolo 8, 35131 Padova, Italy}
\affiliation[c]{INFN, Sezione di Padova, \\Via Marzolo 8, 35131 Padova, Italy}
\affiliation[d]{Physics Faculty, Taras Shevchenko National University of Kyiv, \\64/13 Volodymyrska Street, 01601 Kyiv, Ukraine}
\affiliation[e]{Amherst Center for Fundamental Interactions, Department of Physics, University of Massachusetts,\\1126 Lederle Graduate Research Tower, Amherst, MA 01003-9337, U.S.A.}

\emailAdd{richard.voneckardstein@uni-muenster.de}
\emailAdd{marco.peloso@pd.infn.it}
\emailAdd{kai.schmitz@uni-muenster.de}
\emailAdd{oleksandr.sobol@uni-muenster.de}
\emailAdd{sorbo@physics.umass.edu}


\abstract{Axion inflation coupled to Abelian gauge fields via a Chern--Simons-like term of the form $\phi F\tilde{F}$ represents an attractive inflationary model with a rich phenomenology, including the production of magnetic fields, black holes, gravitational waves, and the matter--antimatter asymmetry. In this work, we focus on a particular regime of axion inflation, the so-called Anber--Sorbo (AS) solution, in which the energy loss in the gauge-field production provides the dominant source of friction for the inflaton motion. We revisit the AS solution and confirm that it is unstable. Contrary to earlier numerical works that attempted to reach the AS solution starting from a regime of weak backreaction, we perform, for the first time, a numerical evolution starting directly from the regime of strong backreaction. Our analysis strongly suggests that, at least as long as one neglects spatial inhomogeneities in the inflaton field, the AS solution has no basin of attraction, not even a very small one that might have been missed in previous numerical studies. Our analysis employs an arsenal of analytical and numerical techniques, some established and some newly introduced, including (1) linear perturbation theory along the lines of Ref.~\cite{Peloso:2022ovc}, (2) the gradient expansion formalism (GEF) developed in Ref.~\cite{Gorbar:2021rlt}, (3) a new linearized version of the GEF, and (4) the standard mode-by-mode approach in momentum space in combination with input from the GEF. All these methods yield consistent results confirming the instability of the AS solution, which renders the dynamics of axion inflation in the strong-backreaction regime even more interesting than previously believed.}


\maketitle


\section{Introduction}
\label{sec:introduction}

Primordial inflation~\cite{Guth:1980zm,Starobinsky:1980te,Linde:1981mu,Albrecht:1982wi} explains some of the puzzles of standard hot big-bang cosmology, and is well supported by observations~\cite{Planck:2018jri,BICEP:2021xfz}. However, the specific particle physics realization of inflation remains unknown, and concrete implementations encounter theoretical challenges. For instance, explaining how the inflaton potential $V \left( \phi \right)$ can maintain flatness against radiative corrections poses a problem. Additionally, in large-field models of inflation, there is the issue of how the inflaton can span a trans-Planckian range. These challenges become even more significant within the swampland program, where bounds on the curvature of scalar potentials and on the excursion of (pseudo)scalar fields~\cite{Palti:2019pca} have been conjectured.

Axion inflation offers a potential solution to the first problem by offering a mechanism that protects the inflaton potential from large radiative corrections~\cite{Freese:1990rb}. Moreover, the most natural coupling of the axion inflaton to gauge fields provides additional elements that might help address both problems~\cite{Anber:2009ua}. It is important to note that the simplest model of axion inflation, called natural inflation~\cite{Freese:1990rb}, has been ruled out by data~\cite{Planck:2018jri,BICEP:2021xfz,Tristram:2021tvh}. However, certain nonminimal realizations, such as the two-axion model of aligned inflation~\cite{Kim:2004rp}, or models of monodromy inflation with a flattened potential~\cite{DAmico:2017cda} have solutions compatible with current data~\cite{Peloso:2015dsa,Copeland:2022lkp}, that will be probed by future CMB experiments, such as CMB-S4~\cite{CMB-S4:2016ple}. In these realizations, the inflaton excursion $\Delta \phi$ is parametrically greater than the axion scale $f$ that characterizes the potentials of the axion(s). Parametrically, this same scale governs the coupling of the inflaton to gauge fields,
\begin{equation}
{\cal L} \supset -\frac{\alpha_\phi}{4 f} \, \phi \, F \, {\tilde F} \;, 
\label{phiFFt}
\end{equation} 
where $F$ is the gauge field strength, ${\tilde F}$ its dual, and $\alpha_\phi$ is a dimensionless coupling that, depending on the specific UV completion of the model, can be expected to be of order one. The motion of the inflaton amplifies one gauge field helicity during inflation~\cite{Anber:2009ua} with a magnitude that is exponentially sensitive to the parameter 
\begin{equation}
\xi \equiv \frac{\alpha_\phi \, \dot{\phi}}{2 f H} \simeq \frac{\alpha_\phi}{\sqrt{2}} \, \frac{\sqrt{\epsilon_V} \, M_\mathrm{P}}{f} \;\;,\;\; \text{standard slow-roll} \;  
\label{xi-def}
\end{equation}
where the dot denotes the derivative with respect to cosmic time; $H$ is the Hubble rate; $\epsilon_V$ is a slow-roll parameter, $\epsilon_V \equiv M_\mathrm{P}^2\left(dV/d\phi \right)^2/\left(2\,V^2\right)$; and $M_\mathrm{P}$ is the reduced Planck scale, related to Newton's constant by $M_\mathrm{P} = (8\pi G_N)^{-1/2}$. In the most straightforward case, $\Delta \phi = \mathcal{O} \left( f \right)$, the axion scale needs to be close to Planckian, and $\xi$ is suppressed by the smallness of $\epsilon_V$. On the contrary, models of aligned natural inflation or of monodromy inflation accommodate an axion scale parametrically smaller than $M_\mathrm{P}$, so the parameter $\xi$ in \eqref{xi-def} can be naturally of order one. 

For $\xi = \mathcal{O} \left( 1 \right)$, the gauge-field amplification induced by its coupling \eqref{phiFFt} to the inflaton can result into a very interesting phenomenology, as the amplified gauge modes can scatter to produce primordial scalar perturbations and gravitational waves (GWs)~\cite{Barnaby:2010vf,Sorbo:2011rz}. Interestingly, this stochastic GW background (SGWB) is circularly polarized, as the coupling \eqref{phiFFt} amplifies only one of the two gauge-field polarizations, and these modes produce one GW polarization much more significantly than the other one. Unfortunately, the strong limits on the gauge-field amplification enforced by the non-observation of the scalar perturbations that they produce prevent this GW signal from being observable at CMB scales.%
\footnote{While in this work we focus on the case of an Abelian gauge field, this problem persists also in the $SU(2)$ case~\cite{Papageorgiou:2018rfx,Papageorgiou:2019ecb} of chromo-natural inflation~\cite{Adshead:2012kp}. The sourced GWs can be observed in specific realizations in which the rolling axion is not the inflaton, so that the production of scalar perturbations is reduced~\cite{Namba:2015gja,Dimastrogiovanni:2016fuu,Thorne:2017jft}. See also Ref.~\cite{Dimastrogiovanni:2023oid} for a non-minimal extension of Ref.~\cite{Adshead:2012kp}.}
The situation might be more interesting at smaller scales. As long as its evolution is described by Eq.~\eqref{xi-def}, the parameter $\xi$ grows during inflation. The exponential sensitivity of the gauge-field production to $\xi$ then results in a GW signal that is naturally much greater at scales smaller than the CMB scale~\cite{Cook:2011hg,Barnaby:2011qe,Domcke:2016bkh}, so that the produced GWs might be observable by a variety of GW observatories~\cite{Garcia-Bellido:2023ser}. Also in this case, one must investigate whether this potential GW signature can take place without a simultaneous overproduction of scalar perturbations, which would lead to too many primordial black holes (PBHs)~\cite{Linde:2012bt,Bugaev:2013fya,Garcia-Bellido:2016dkw,Garcia-Bellido:2017aan,Ozsoy:2023ryl}. For reasons that we discuss in the remainder of this Introduction, we still do not have a reliable answer to this question. 

When the parameter $\xi$ is sufficiently large, the amplified gauge fields significantly backreact on the background dynamics. The backreaction occurs via an $\alpha_\phi \left\langle \bm{E} \cdot \bm{B} \right\rangle/f$ term in the equation of motion (EOM) of the inflaton and via the gauge-field energy density $\propto \left\langle \bm{E} \cdot \bm{E} \right\rangle + \left\langle \bm{B} \cdot \bm{B} \right\rangle$ in the Friedmann equation for the Hubble rate.%
\footnote{We use electromagnetic notation for simplicity, even though we are not necessarily assuming that the gauge field coupled to the inflaton is the Standard Model photon or the hypercharge gauge boson.}
The former effect is typically more relevant, as it is easier to impact the dynamics of the inflaton, which is slow-roll suppressed, than that of the scale factor. Anber and Sorbo (AS) investigated this model for the case where the background evolution is always in a regime of strong backreaction, in which the dissipation due to the gauge-field production provides the dominant source of friction (over the standard $3 H \dot{\phi}$ Hubble friction term) for the inflaton motion~\cite{Anber:2009ua}. Ref.~\cite{Barnaby:2011qe} considered instead the case where the evolution is in a regime of weak backreaction (in which the gauge-field amplification negligibly impacts the background dynamics) at the time when the CMB modes exit the horizon, followed by a smooth  transition to the regime of strong backreaction, causing the generation of a visible GW signal at smaller scales. In these and in several successive works, the strong-backreaction AS regime was studied under the assumption that the inflaton speed, and hence the parameter $\xi \left( t \right)$, follow a slow and monotonic steady-state evolution, characterized by the friction due to gauge-field amplification perfectly balancing the gradient force from the inflaton potential at all times. This is the typical behavior that is expected in a realization of warm inflation~\cite{Berera:1995ie}. 

In the past few years, this system has been studied with different numerical schemes of increasing precision and sophistication. Refs.~\cite{Cheng:2015oqa,Notari:2016npn,DallAgata:2019yrr,Garcia-Bellido:2017aan,Cado:2022pxk} numerically integrated the evolution for the case of a homogeneous inflaton coupled to a large set of gauge-field modes $A_\lambda\left(\tau,k\right)$. The authors of Ref.~\cite{Domcke:2020zez} adopted a recursive integration approach to study the same system of equations. They initially integrated the equations for the gauge modes using an ``external'' inflaton and scale factor evolution, where the backreaction of the gauge modes is neglected. Next, they employed these gauge-mode solutions as ``external functions'' for the backreaction in the evolution equations for the inflaton and scale factor. In this way, they obtained improved solutions for these two quantities from which they could then obtain improved solutions for the gauge modes. By iteratively updating this procedure, they achieved convergence towards a consistent solution encompassing all these quantities. Refs.~\cite{Gorbar:2021rlt,Gorbar:2021zlr,Durrer:2023rhc} took a different approach~\cite{Sobol:2019xls} by considering a set of equations for the two-point correlators of the ``electric'' and ``magnetic'' combinations. These equations involve other two-point correlators that include spatial derivatives, such as $\left\langle \bm{E} \cdot \operatorname{rot}^n \bm{B} \right\rangle$. By constructing a hierarchy of equations involving correlators with an increasing number of spatial derivatives, one can numerically solve them after truncating the hierarchy at a certain order. This allows for the computation of the correlators and provides a systematic method for studying the dynamics of the system. This gradient expansion formalism (GEF) is extended to a computation of linear perturbations in this work. 

While all these studies assumed a homogeneous inflaton, the equations for a gauge field coupled to a spacetime-dependent inflaton were studied in Refs.~\cite{Caravano:2022epk,Figueroa:2023oxc} on the lattice.
\footnote{While this discussion is focused on the inflationary evolution, lattice simulations of reheating at the end of inflation can be found in Refs.~\cite{Adshead:2015pva,Adshead:2016iae,Figueroa:2017qmv,Adshead:2018doq,Cuissa:2018oiw,Adshead:2019lbr,Adshead:2019igv,Figueroa:2021yhd}.} 
The lattice simulations conducted in Ref.~\cite{Caravano:2022epk} yielded results for the inflaton power spectrum and bispectrum in the weak-backreaction regime, which exhibited excellent agreement with the analytical computations of Refs.~\cite{Barnaby:2010vf,Barnaby:2011vw}. However, the findings from the numerical studies~\cite{Cheng:2015oqa,Notari:2016npn,DallAgata:2019yrr,Garcia-Bellido:2017aan,Cado:2022pxk,Caravano:2022epk,Figueroa:2023oxc,Domcke:2020zez,Gorbar:2021rlt,Gorbar:2021zlr,Durrer:2023rhc,Sobol:2019xls} of the strong-backreaction regime contradicted the analytical expectations. It was discovered that the evolution of the inflaton does not occur in a steady-state regime; instead, the parameter $\xi \left( t \right)$ undergoes large oscillations, with a period of approximately $\sim 5$ $e$-folds, around the steady-state analytical solution. Ref.~\cite{Domcke:2020zez} showed that these oscillations can be attributed to a memory effect. Specifically, a gauge-field mode begins to undergo amplification when its reduced wavelength is approximately a factor of $2\xi$ smaller than the Hubble horizon, $\lambdabar \equiv \lambda/\left(2\pi\right) \simeq H^{-1}/\left(2\xi\right)$. The amplification ceases shortly after the mode crosses the horizon at $\lambdabar = H^{-1}$, and the energy of the mode is subsequently diluted by the expansion of the Universe. This introduces a sensitivity of the backreaction at a given time $t$ to the evolution of the system during the previous few $e$-folds. This sensitivity gives rise to an oscillatory behavior in the derivative of the inflaton field, $\dot{\phi}(t)$. These oscillations aim to ``adjust'' the cumulative effect of the gauge fields amplified during the preceding few $e$-folds, aligning it with the slope of the potential at the specific moment in question.\footnote{To avoid this problem, and generate a stable steady-state dissipative regime, Ref.~\cite{Creminelli:2023aly} recently provided a construction, based on scalar field interactions, that can generate an ``instantaneous'' sensitivity of the backreaction to the particle production.}

This interpretation was confirmed by the analytical study in Ref.~\cite{Peloso:2022ovc}, which solved the linearized set of equations for the homogeneous inflaton perturbation, $\delta\sm \phi$, and gauge-mode perturbations about an AS solution with constant $H$ and $\dot{\phi}$. The equation for the gauge-mode perturbations can be formally solved in terms of a Green function (constructed from the gauge modes of the unperturbed AS solution) and the inflaton perturbation. These formal solutions were then substituted back into the equation for $\delta\sm \phi$, resulting in an integro-differential equation where the memory effects are encoded in the kernel of the integral. Through suitable simplifications, Ref.~\cite{Peloso:2022ovc} reduced this equation to an algebraic equation, and the roots of this equation were then determined numerically. In addition to providing a relationship between the growth and period of the oscillations and the model parameters, this analysis differs from the previous ones, as it addresses the instability of the AS solution assuming it as an initial condition rather than considering an evolution that started in the weak-backreaction regime and that was expected to evolve into the AS solution. This explored the stability of the AS solution itself,  excluding the possibility that it may only have a small basin of attraction that was not reached by the existing studies. 

The findings of Ref.~\cite{Peloso:2022ovc} confirmed the instability previously observed in numerical studies. These analytical results were, however, obtained with considerable simplifications (particularly, with respect to the form of the Green function), and it is therefore important to confirm them with a numerical study that, contrary to the existing ones, assumes the AS solution as initial condition. This is the goal of the present work. We do this with two independent methods, both based on the GEF. Firstly, we consider the linearized system of perturbations considered in Ref.~\cite{Peloso:2022ovc} and rewrite them in GEF language. We do not apply any analytical approximation to these equations, but we rather solve them numerically, using the same general ansatz for $\delta\sm \phi$ adopted in Ref.~\cite{Peloso:2022ovc}. Our solutions provide more accurate results compared to Ref.~\cite{Peloso:2022ovc}, albeit at the cost of a less direct connection with the model parameters. These improved results confirm the presence of unstable modes observed in Ref.~\cite{Peloso:2022ovc}, yielding more precise values for the growth and period of the oscillations. Secondly, we employ the GEF equations to conduct a full numerical study, starting from the AS solution as initial condition. This approach allows us to consistently incorporate variations in the Hubble rate and the inflaton speed, which occur in concrete inflationary models but were neglected in the analytical computations of Ref.~\cite{Peloso:2022ovc} and our first method. Importantly, this represents the first numerical evidence of the instability of the AS solution in the strong-backreaction regime, assuming its validity at the beginning of the evolution. 

The work is organized as follows. In Sec.~\ref{sec:toolkit}, we review the basics of the axion inflation model that we are going to be interested in, presenting all the equations and tools that are generally used to study the dynamics of the inflaton and gauge fields in this model. We also precisely define the AS solution and discuss in which part of parameter space we expect this solution to be realized. In Sec.~\ref{sec:const_back}, we consider the simplified case of a constant Hubble rate and a constant inflaton dragging force, such that the AS solution corresponds to a constant inflaton velocity. After discussing two methods that allow us to study the stability of the AS solution in the linear perturbation regime (the method in Ref.~\cite{Peloso:2022ovc} as well as the linearized GEF), we present the numerical results showing the spectrum of Lyapunov exponents of the linear system as well as the survival time of the AS solution and the late-time behavior of the exact solution in the nonlinear regime. In Sec.~\ref{sec:realistic}, we consider the case of a realistic inflationary model where the Hubble rate is now consistently determined by the Friedmann equation and analyze the stability of the AS solution in this case. Sec.~\ref{sec:conclusion} is devoted to our conclusions. In Appendix~\ref{app:bilinears}, we list some auxiliary formulas for the bilinear functions in the GEF. In Appendix~\ref{app:survival-time}, we explain how we impose the initial conditions for the GEF system for the purposes of the analysis in Sec.~\ref{sec:const_back}, and we study the dependence of the survival time of the AS solution on the axion inflation model parameters and on the initial conditions. Finally, in Appendix~\ref{app:self-correction}, we give more details on a novel self-correction procedure that we employ in our GEF computations and which allows us to extend the applicability of the GEF to later times. Throughout the work, we use natural units and set $\hbar=c=1$; then, the reduced Planck mass reads $M_{\mathrm{P}}= 2.43\times 10^{18}\,\text{GeV}$. We assume that the Universe is described by a spatially flat Friedmann--Lema\^{i}tre--Robertson--Walker (FLRW) metric with line element (in terms of cosmic time $t$ and conformal time $\tau$)
\begin{equation}
    ds^2=g_{\mu\nu}dx^\mu dx^\nu=-dt^2+a^2(t)d\bm{x}^2=a^2(\tau)\left[-d\tau^2+d\bm{x}^2\right] \,.
\end{equation}


\section{Model and tools}
\label{sec:toolkit}


\subsection{Axion inflation}

In the axion inflation model, the pseudoscalar inflaton field $\phi$ couples to a $U(1)$ gauge field $A_\mu$ via a Chern--Simons-like term. The corresponding  action has the following form:
\begin{equation}
\label{action-1}
    S[\phi,\,A_\mu]=\int d^4 x\sqrt{-g}\left[-\frac{1}{2}g^{\mu\nu}\partial_\mu\phi\partial_\nu\phi-V(\phi)-\frac{1}{4}F_{\mu\nu}F^{\mu\nu}-\frac{\beta\,\phi}{4M_{\mathrm{P}}}F_{\mu\nu}\tilde{F}^{\mu\nu}\right],
\end{equation}
where $V(\phi)$ is the inflaton potential; $F_{\mu\nu}=\partial_\mu A_\nu-\partial_\nu A_\mu$ is the gauge-field tensor; $\tilde{F}^{\mu\nu}=\varepsilon^{\mu\nu\alpha\beta}F_{\alpha\beta}/(2\sqrt{-g})$ is the associated dual tensor with totally antisymmetric Levi-Civita symbol, $\varepsilon^{\mu\nu\alpha\beta}$ with $\varepsilon^{0123} = 1$; and $\beta \equiv \alpha_\phi \, M_{\mathrm{P}} / f$ is a dimensionless axion--vector coupling constant. From Eq.~\eqref{action-1}, we can compute the inflaton and gauge-field EOMs,
\begin{align}
\label{KGF}
\frac{1}{\sqrt{-g}}\partial_{\mu}\left[\sqrt{-g}\,g^{\mu\nu} \partial_{\nu}\phi \right]-\frac{dV}{d\phi}-\frac{\beta}{4M_\mathrm{P}}\,F_{\mu\nu}\tilde{F}^{\mu\nu} & = 0 \,, \\
\label{Maxwell-cov}
\frac{1}{\sqrt{-g}}\partial_{\mu}\left[\sqrt{-g}\,F^{\mu\nu} \right]+ \frac{\beta}{M_\mathrm{P}}\,\tilde{F}^{\mu\nu}\partial_{\mu}\phi & = 0 \,,
\end{align}
where the latter equation is supplemented by the Bianchi identity for the dual tensor,
\begin{equation}
    \label{Maxwell-Bianchi}
    \frac{1}{\sqrt{-g}}\,\partial_{\mu}\left[\sqrt{-g}\,\tilde{F}^{\mu\nu} \right]=0 \,.
\end{equation}

In temporal gauge, the gauge field is written as $A_\mu=(0,\,\bm{A})$. Then, the three-vectors of the physical electric field $\bm{E}$ and magnetic field $\bm{B}$ in the comoving frame are defined as
\begin{equation}
\bm{E}=-\frac{1}{a}\frac{\partial \bm{A}}{\partial t} \,, \qquad \bm{B}=\frac{1}{a^2} \operatorname{rot}\bm{A} \,.
\end{equation}
Correspondingly, the components of the gauge-field tensor and its dual are expressed in terms of the components of $\bm{E}$ and $\bm{B}$ in the following way:
\begin{equation}
F_{0i}=-aE^{i} \,, \qquad F_{ij}=a^2 \varepsilon_{ijk}B^{k} \,, \qquad \tilde{F}_{0i}=-aB^{i} \,, \qquad \tilde{F}_{ij}=-a^2 \varepsilon_{ijk} E^{k} \,,
\end{equation}
where $\varepsilon_{ijk}$ is the totally antisymmetric Levi-Civita symbol in three spatial dimensions.

The energy--momentum tensor following from the action \eqref{action-1} reads
\begin{equation}
\label{T}
T^\mu_\nu=-\partial^{\mu}\phi\,\partial_{\nu}\phi - F^{\mu\lambda}F_{\nu\lambda}+\delta^\mu_{\nu}\left[\frac{1}{2}\partial_{\alpha}\phi\,\partial^{\alpha}\phi+V(\phi)+\frac{1}{4}F_{\alpha\beta}F^{\alpha\beta}\right] \,,
\end{equation}
which, assuming a spatially homogeneous inflaton field, yields an energy density
\begin{equation}
\label{rho}
\rho=\langle T_{0}^{0}\rangle=\frac{1}{2}\dot{\phi}^{2}+V(\phi) + \frac{1}{2}\langle \bm{E}^{2}+\bm{B}^{2} \rangle \,,
\end{equation}
where the angle brackets around $T_0^0$ and $ \bm{E}^{2}+\bm{B}^{2}$ denote the expectation value during inflation. The energy density determines the Hubble expansion rate $H$ through the Friedmann equation,
\begin{equation}
\label{Friedmann}
    H^2=\frac{\rho}{3M_\mathrm{P}^2} \,.
\end{equation}
Finally, we rewrite Eqs.~\eqref{KGF}--\eqref{Maxwell-Bianchi} in three-vector form, 
\begin{align}
\label{KGF-2}
\ddot{\phi}+3H\dot{\phi}+V'(\phi) & = \frac{\beta}{M_\mathrm{P}}\langle \bm{E}\cdot\bm{B} \rangle \,, \\
\label{Maxwell_1}
\dot{\bm{E}}+2 H \bm{E}-\frac{1}{a} \operatorname{rot} \bm{B} + \frac{\beta}{M_\mathrm{P}}\,\dot{\phi}\,\bm{B} & = 0 \,, \\
\label{Maxwell_2}
\dot{\bm{B}}+2 H \bm{B}+\frac{1}{a} \operatorname{rot} \bm{E}& = 0 \,, \vphantom{\frac{\beta}{M_\mathrm{P}}} \\
\label{Maxwell_3}
\operatorname{div} \bm{E}=0, \qquad \operatorname{div} \bm{B} & = 0 \,. \vphantom{\frac{\beta}{M_\mathrm{P}}}
\end{align}
The system of equations \eqref{Friedmann}--\eqref{Maxwell_3} is a complete set of equations governing the joint evolution of the homogeneous inflaton field, scale factor, and gauge fields in position space during axion inflation.

Let us now switch to momentum space and consider the quantized gauge field
\begin{equation}
\label{quantized_A}
\hat{\bm{A}}(t,\bm{x})=\int\frac{d^{3}\bm{k}}{(2\pi)^{3/2}}\sum_{\lambda=\pm}\left[\bm{\epsilon}^{\lambda}(\bm{k})\hat{a}_{\bm{k},\lambda}A_{\lambda}(t,k)e^{i\bm{k}\cdot\bm{x}}+\bm{\epsilon}^{\lambda *}(\bm{k})\hat{a}_{\bm{k},\lambda}^{\dagger}A_{\lambda}^{*}(t,k)e^{-i\bm{k}\cdot\bm{x}} \right] \,,
\end{equation}
with mode functions $A_{\lambda}(t,k)$, polarization three-vectors $\bm{\epsilon}^{\lambda}(\bm{k})$, and annihilation (creation) operators $\hat{a}_{\bm{k},\lambda}$ ($\hat{a}^{\dagger}_{\bm{k},\lambda}$) for electromagnetic modes with momentum $\bm{k}$ and circular polarization $\lambda=\pm$, and $k=|\bm{k}|$. The polarization vectors satisfy the relations
\begin{equation}
\bm{k}\cdot\bm{\epsilon}^{\lambda}(\bm{k})=0,\quad \bm{\epsilon}^{\lambda *}(\bm{k})=\bm{\epsilon}^{-\lambda}(\bm{k}), \quad [i\bm{k}\times\bm{\epsilon}^{\lambda}(\bm{k})]=\lambda k \bm{\epsilon}^{\lambda}(\bm{k}), \quad \bm{\epsilon}^{\lambda *}(\bm{k})\cdot\bm{\epsilon}^{\lambda'}(\bm{k})=\delta^{\lambda\lambda'} \,,
\end{equation}
Note that the first relation is equivalent to choosing Coulomb gauge in addition to temporal gauge, which we can impose since $\operatorname{div} \bm{E}=0$. The creation and annihilation operators obey the canonical commutation relations
\begin{equation}
[\hat{a}_{\bm{k},\lambda},\,\hat{a}^{\dagger}_{\bm{k}',\lambda'}]=\delta_{\lambda\lambda'}\delta^{(3)}(\bm{k}-\bm{k}')\,.
\end{equation}

For the vector potential in Eq.~\eqref{quantized_A}, the Maxwell equations~\eqref{Maxwell_2}--\eqref{Maxwell_3} are identically satisfied, while Eq.~\eqref{Maxwell_1} leads to an EOM for the mode functions,
\begin{equation}
\label{A_1}
\ddot{A}_{\lambda}(t,k)+H\dot{A}_{\lambda}(t, k)+\left[\frac{k^{2}}{a^{2}}-\lambda\frac{k}{a}\frac{\beta}{M_{\mathrm{P}}}\dot{\phi}\right] A_{\lambda}(t,k) =0 \,,
\end{equation}
which takes a slightly simpler form when written in conformal time $\tau=\int^{t}dt'/a(t')$,
\begin{equation}
    \label{A_2}
    \frac{\partial^2 A_\lambda(\tau, k)}{\partial\tau^2}+\left[k^2-\lambda k\frac{\beta}{M_{\mathrm{P}}}\frac{d\phi}{d\tau}\right] A_\lambda(\tau, k)=0 \,.
\end{equation}
Deep inside the horizon, $k\tau\to-\infty$, the first term inside the square brackets dominates, and Eq.~\eqref{A_2} takes the form of a simple harmonic-oscillator equation. This allows us to impose the Bunch--Davies boundary condition~\cite{Bunch:1978yq} in the asymptotic past, which amounts to selecting the flat-space positive-frequency solutions for modes deep inside the horizon,
\begin{equation}
\label{BD-condition}
    A_\lambda(\tau, k)\simeq\frac{1}{\sqrt{2k}}e^{-ik\tau},\qquad k\tau\to -\infty\,.
\end{equation}

Next, we use Eq.~\eqref{quantized_A} to compute the vacuum expectation value of $\bm{E}\cdot \bm{B}$ on the right-hand side of Eq.~\eqref{KGF-2}. Then, in conformal time, the Klein--Gordon equation reads
\begin{equation}
    \label{KGF-sp}
    \frac{d^2\phi}{d\tau^2}+2aH\frac{d\phi}{d\tau}+a^2 V'(\phi)=-\frac{\beta}{4\pi^2 a^2 M_\mathrm{P}}\int\limits_0^{k_{\mathrm{h}}}dk\,k^3 \frac{\partial}{\partial \tau}\left[\big|A_{+}(\tau,k)\big|^2-\big|A_{-}(\tau,k)\big|^2\right]\,.
\end{equation}
Finally, we express the gauge-field energy density in terms of the mode functions. Then, the Friedmann equation for the Hubble rate $H =(da/d\tau)/a^2$ takes the form
\begin{equation}
\label{Friedmann-2}
H^2 = \frac{1}{3M_{\mathrm{P}}^2}\bigg\{\frac{1}{2a^2}\left(\frac{d\phi}{d\tau}\right)^2+V(\phi)+\frac{1}{4\pi^2 a^4}\int\limits_0^{k_{\mathrm{h}}}dk\,\sum\limits_{\lambda=\pm}\left[k^4\big|A_{\lambda}\big|^2 +k^2\Big|\frac{\partial A_{\lambda}}{\partial\tau}\Big|^2\right]\bigg\} \,.
\end{equation}
Here, the momentum scale $k_{\mathrm{h}}$ in Eqs.~\eqref{KGF-sp} and \eqref{Friedmann-2} denotes a finite upper integration limit. This cutoff is necessary to separate the physically relevant gauge-field modes, which undergo enhancement due to the axial coupling, from pure vacuum fluctuations. We choose $k_{\mathrm{h}}$  such that the bracket in the mode equation~\eqref{A_2} vanishes for one polarization state,
\begin{equation}
k_{\mathrm{h}} = \left|\frac{\beta}{M_{\mathrm{P}}}\frac{d\phi}{d\tau}\right| \,.
\label{cutoff}
\end{equation}
In this way, the integrals in Eqs.~\eqref{KGF-sp} and \eqref{Friedmann-2} include all modes for which one polarization has experienced the tachyonic instability. In summary, Eqs.~\eqref{A_2}--\eqref{Friedmann-2} represent the full set of equations describing the dynamics of axion inflation in momentum space.


\subsection{Gradient expansion formalism}
\label{subsec:GEF}

An alternative way to treat axion inflation in position space is the gradient expansion formalism (GEF)~\cite{Gorbar:2021rlt}. Let us introduce the following set of bilinear gauge-field functions:
\begin{align}
\label{E_n}
\mathcal{E}^{(n)} & \equiv \frac{1}{a^{n}}\left\langle \bm{E}\cdot\operatorname{rot}^{n} \bm{E}  \right\rangle \,, \\
\label{G_n}
\mathcal{G}^{(n)} & \equiv -\frac{1}{2a^{n}}\left\langle \bm{E}\cdot \operatorname{rot}^{n} \bm{B} + \operatorname{rot}^{n}\bm{B} \cdot \bm{E}\right\rangle \,, \\
\label{B_n}
\mathcal{B}^{(n)} & \equiv \frac{1}{a^{n}}\left\langle \bm{B}\cdot \operatorname{rot}^{n} \bm{B}  \right\rangle \,,
\end{align}
and recast the Maxwell equations~\eqref{Maxwell_1} and \eqref{Maxwell_2} as an infinite tower of equations,
\begin{align}
\label{dot_E_n}
& \dot{\mathcal{E}}^{(n)} + (n+4)H\,	\mathcal{E}^{(n)} - \frac{2\beta}{M_\mathrm{P}}\dot{\phi}\,\mathcal{G}^{(n)} + 2\mathcal{G}^{(n+1)} = S_{\mathcal{E}} \,, \\
\label{dot_G_n}
& \dot{\mathcal{G}}^{(n)} +(n+4)H\, \mathcal{G}^{(n)}- \frac{\beta}{M_\mathrm{P}}\dot{\phi}\,\mathcal{B}^{(n)} 
-\mathcal{E}^{(n+1)}+\mathcal{B}^{(n+1)} = S_{\mathcal{G}} \,, \\
\label{dot_B_n}
& \dot{\mathcal{B}}^{(n)} + (n+4)H\,\mathcal{B}^{(n)}-2\mathcal{G}^{(n+1)} = S_{\mathcal{B}} \,. \vphantom{\frac{\beta}{M_{\mathrm{P}}}}
\end{align}
Here, the source terms on the right-hand side of the equations denote boundary terms that account for the fact that the number of physically relevant modes changes in time during inflation, as a consequence of the fact that the cutoff scale in Eq.~\eqref{cutoff} changes with time. These boundary terms were derived in Ref.~\cite{Gorbar:2021rlt} and are given by the expressions
\begin{align}
\label{SE}
S_{\mathcal{E}} & = \frac{d \ln k_{\mathrm{h}}(t)}{d t}\frac{1}{4\pi^{2}}\left(\frac{k_{\mathrm{h}}(t)}{a(t)}\right)^{n+4}\sum_{\lambda=\pm 1}\lambda^{n}E_\lambda\left(\xi(t)\right) \,, \\
\label{SG}
S_{\mathcal{G}} & = \frac{d \ln k_{\mathrm{h}}(t)}{d t}\frac{1}{4\pi^{2}}\left(\frac{k_{\mathrm{h}}(t)}{a(t)}\right)^{n+4}\sum_{\lambda=\pm 1}\lambda^{n+1}G_\lambda\left(\xi(t)\right) \,, \\
\label{SB}
S_{\mathcal{B}} & = \frac{d \ln k_{\mathrm{h}}(t)}{d t}\frac{1}{4\pi^{2}}\left(\frac{k_{\mathrm{h}}(t)}{a(t)}\right)^{n+4}\sum_{\lambda=\pm 1}\lambda^{n}B_\lambda\left(\xi(t)\right) \,,
\end{align}
where the auxiliary functions $E_\lambda$, $G_\lambda$ and $B_\lambda$ are given in terms of Whittaker functions,
\begin{align}
\label{E_lambda}
E_\lambda(\xi) & = \frac{e^{\pi\lambda \xi}}{4\xi^2}\left|\left(2i|\xi| -i\lambda \xi\right)W_{-i\lambda\xi,\frac{1}{2}}(-4i |\xi|)+W_{1-i\lambda\xi,\frac{1}{2}}(-4i|\xi|)\right|^{2} \,, \\
\label{G_lambda}
G_\lambda(\xi) & =\frac{e^{\pi\lambda \xi}}{|2\xi|}\operatorname{Re}\left[W_{i\lambda \xi,\frac{1}{2}}(4i|\xi|)\, W_{1-i\lambda\xi,\frac{1}{2}}(-4i|\xi|)\right] \,, \\
\label{B_lambda}
B_\lambda(\xi) & = e^{\pi\lambda \xi}\,\left|W_{-i\lambda\xi,\frac{1}{2}}(-4i|\xi|) \right|^{2} \,,
\end{align}
and where the gauge-field production parameter $\xi$ was already introduced in Eq.~\eqref{xi-def},
\begin{equation}
\label{xi}
    \xi \equiv \frac{\beta \dot{\phi}}{2HM_{\mathrm{P}}} \,.
\end{equation}
The quantity $k_{\mathrm{h}}(t)$ in Eqs.~\eqref{SE}--\eqref{SB} is the wavenumber of the highest-momentum mode that has ever become tachyonically unstable during the entire evolution of the system leading up to the moment of time $t$,
\begin{equation}
\label{k-h-final}
k_\mathrm{h}(t)\equiv\max\limits_{t'\leq t}\left\{a(t')\left|\frac{\beta\dot{\phi}(t')}{M_\mathrm{P}}\right|\right\} \,.
\end{equation}
Note that this definition slightly deviates from the momentum scale $k_\mathrm{h}$ defined in Eq.~\eqref{cutoff}; in fact, $k_\mathrm{h}(t)$ in Eq.~\eqref{k-h-final} represents an improved version of $k_{\mathrm{h}}$ in the sense that it accounts for the fact that the evolution of the right hand side of Eq.~\eqref{cutoff} is not monotonic when the inflaton velocity oscillates. Employing $k_{\mathrm{h}}$ in Eq.~\eqref{cutoff} as the upper integration boundary, therefore, leads to situations where modes that already experienced the tachyonic instability, and which should thus be regarded as physically excited, fall into the region above the UV cutoff and are hence removed from the momentum integrals. With the improved definition, a mode is accounted for at all times after it has become unstable for the first time. In the following, we will exclusively work with the definition in Eq.~\eqref{k-h-final}.

Finally, in order to solve the system of equations numerically, we need to truncate it at some finite order $n_{\mathrm{cut}}$. This can be done by expressing the quantities of order $(n_{\mathrm{cut}}+1)$ through expressions of lower order. One of the simplest ways to truncate the system was discussed in Ref.~\cite{Gorbar:2021rlt} and it is based on the following relation:
\begin{equation}
\label{truncation}
    \mathcal{E}^{(n_{\mathrm{cut}}+1)} = \left(\frac{k_{\mathrm{h}}(t)}{a(t)}\right)^{2}\mathcal{E}^{(n_{\mathrm{cut}}-1)} \;, 
\end{equation}
and analogously for $\mathcal{G}^{(n_{\mathrm{cut}}+1)}$ and $\mathcal{B}^{(n_{\mathrm{cut}}+1)}$. This truncation condition can be justified by the following consideration. For sufficiently large order $n$, the spectrum of bilinear quantities $\mathcal{E}^{(n)}$, $\mathcal{G}^{(n)}$, and $\mathcal{B}^{(n)}$ is always blue and, therefore, the dominant contribution stems from modes with momenta $k\simeq k_{\mathrm{h}}$. Then, the mean value theorem for the integral over momentum leads to the condition~\eqref{truncation}~\cite{Gorbar:2021rlt}.
Knowing the behavior of the spectrum near the cutoff momentum $k_{\mathrm{h}}$, one could estimate the error of the truncation introduced by the truncation condition. In practice, however, the truncation order $n_{\mathrm{cut}}$ is chosen in such a way that increasing $n_{\mathrm{cut}}$ further does not have an impact on the solution.


\subsection{Anber--Sorbo solution}

If the gauge field is absent (or, at least, sufficiently weak such that its contributions to the Friedmann and Klein--Gordon equations can be neglected) and the potential $V(\phi)$ is sufficiently flat, the inflaton $\phi$ follows the slow-roll attractor solution where the Hubble friction term is almost exactly compensated by the gradient force from the inflaton potential, 
\begin{equation}
    \label{slow-roll}
    3H\dot{\phi}+V'(\phi)\simeq 0,\qquad 3H^2 M_\mathrm{P}^2\simeq V(\phi).
\end{equation}
This describes the usual case of slow-roll inflation, which is a true attractor solution: any initial deviation quickly tends to zero, leading the system into the slow-roll regime.

The idea of Anber and Sorbo in Ref.~\cite{Anber:2009ua} was to realize the inflationary stage even with a steep potential $V(\phi)$ (for which the standard slow-roll regime is impossible) due to the backreaction from the produced gauge fields.
In this case, the potential-gradient force is compensated by the gauge-field friction term on the right-hand side of the Klein--Gordon equation,
\begin{equation}
\label{AS}
V'(\phi)\simeq \frac{\beta}{M_\mathrm{P}}\langle\bm{E}\cdot\bm{B}\rangle.
\end{equation}
However, this solution is now known to be unstable, as discussed in detail in the Introduction. A first analytical computation demonstrating the instability of the AS solution was recently presented by two of us (M.\,P.\ and L.\,S.) in Ref.~\cite{Peloso:2022ovc}. In the present paper, we shall substantiate this computation by a comprehensive numerical analysis that will allow us to achieve two results: (i) precisely determine the growth rate of instabilities around the AS solution and hence (ii) measure the survival time of the AS solution, which we define as the time when the relative deviation from the AS solution becomes of $\mathcal{O}\left(1\right)$.

First of all, let us give a precise definition of the AS solution that we will use throughout the paper. Let us assume for a moment that the AS solution is indeed an attractor solution, as initially proposed in Ref.~\cite{Anber:2009ua} and subsequently assumed as a working hypothesis in many papers in the literature. Then, under this assumption, we expect the system to slowly evolve in time because of the strong friction coming from the gauge-field backreaction. In this regime, it is natural to assume that the value of the gauge field at a given moment of time is determined by the inflaton velocity at the same moment of time, \textit{i.e.}, there is no retardation between the inflaton evolution and the gauge-field response. In this case, the Klein--Gordon equation~\eqref{KGF-2} represents a closed equation for the inflaton field,
\begin{equation}
\label{KGF-AS}
\ddot{\phi}(t)+3H(t)\dot{\phi}(t)+V'(\phi)=\frac{\beta}{M_\mathrm{P}}\langle \bm{E}\cdot\bm{B} \rangle [\dot{\phi}(t)] \,,
\end{equation}
where we now emphasize that $\langle \bm{E}\cdot\bm{B} \rangle$ is a functional of the inflaton velocity $\dot{\phi}$ and where the Hubble parameter follows from the Friedmann equation in the usual way,
\begin{equation}
H^2\left(t\right) = \frac{1}{3M_\mathrm{P}^2}\left[\frac{1}{2}\,\dot{\phi}^2(t) + V\left(\phi\right)+\frac{1}{2}\left(\langle\bm{E}^2\rangle[\dot{\phi}(t)]+\langle\bm{B}^2\rangle[\dot{\phi}(t)]\right)\right] \,.
\end{equation}
In order to find the explicit form of the functional dependence of the energy densities and the Chern--Pontryagin density on $\dot{\phi}$, we assume again that the inflaton is rolling slowly and the Universe is expanding quasi-exponentially. Therefore, on the timescale of a few $e$-folds, one can assume that the Hubble parameter $H$ and the inflaton velocity $\dot{\phi}$ are approximately constant. This significantly simplifies the mode equation~\eqref{A_2} in conformal time,
\begin{equation}
\label{A-const-xi-H}
\frac{\partial^2 A_\lambda(\tau, k)}{\partial \tau^2}+\left[k^2+2\lambda k\frac{\xi}{\tau}\right] A_\lambda(\tau, k)=0 \,,
\end{equation}
where we used the relation $a(\tau)=-1/(H\tau)$ for de Sitter space and where the parameter $\xi$ introduced in Eq.~\eqref{xi} is now constant (for definiteness, let us assume $\xi>0$). In this case, Eq.~\eqref{A-const-xi-H} has an exact solution in terms of Whittaker functions. The Bunch--Davies boundary condition~\eqref{BD-condition} allows to extract a unique solution of the form
\begin{equation}
\label{A-Whittaker}
A_\lambda\left(\tau, k\right) = \frac{e^{\lambda\pi\xi/2}}{\sqrt{2k}} \,W_{-i\lambda\xi,\frac{1}{2}}\left(2ik\tau\right) \,,
\end{equation}
where $W_{\kappa,\mu}(z)$ is the Whittaker $W$ function. Eq.~\eqref{A-const-xi-H} implies that negative-helicity modes $A_{-}$ are not enhanced, because the expression in square brackets is always positive. However, positive-helicity modes $A_{+}$ experience a tachyonic instability when $k<2\xi\,aH$, which leads to their exponential amplification. This can be directly seen from the following approximate expression for $A_{+}$ in terms of elementary functions, which is valid for $x\ll 2\xi$~\cite{Anber:2009ua}:
\begin{equation}
\label{A-plus-approx}
    A_{+}(\tau,k)\simeq \frac{1}{\sqrt{2k}}\bigg[\left(\frac{x}{2\xi}\right)^{1/4}e^{\pi\xi-2\sqrt{2\xi x}}+\frac{i}{2}\left(\frac{x}{2\xi}\right)^{1/4}e^{-\pi\xi+2\sqrt{2\xi x}}\bigg],\qquad x\equiv -k\tau \,.
\end{equation}

Now, substituting Eq.~\eqref{A-Whittaker} into the expressions for the Chern--Pontryagin density and energy densities, we find 
\begin{equation}
\label{e0b0g0}
\langle\bm{E}^2\rangle=H^4 e_0(\xi) \,, \qquad \langle\bm{E}\cdot\bm{B} \rangle=-H^4 g_0(\xi) \,, \qquad \langle\bm{B}^2 \rangle=H^4 b_0(\xi) \,,
\end{equation}
where the functions $e_0$, $g_0$, and $b_0$ are given in the form of integrals of the Whittaker functions \eqref{IC-En}--\eqref{IC-Bn} in Appendix~\ref{app:bilinears}. Using Eq.~\eqref{A-plus-approx}, we can approximately write
\begin{equation}
\label{e0b0g0-approx}
    e_0(\xi)\approx \frac{6!}{2^{18}\pi^2}\frac{e^{2\pi\xi}}{\xi^3}, \qquad  g_0(\xi)\approx \frac{7!}{2^{21}\pi^2}\frac{e^{2\pi\xi}}{\xi^4}, \qquad b_0(\xi)\approx \frac{8!}{2^{24}\pi^2}\frac{e^{2\pi\xi}}{\xi^5}.
\end{equation}

Therefore, we now define the AS solution as a solution of the system of equations
\begin{align}
\label{KGF-AS-2}
& \ddot{\phi}+3H\dot{\phi}+V'(\phi)=-\frac{\beta}{M_\mathrm{P}}H^4 g_0\left(\tfrac{\beta \dot{\phi}}{2HM_{\mathrm{P}}}\right) \,, \\
\label{Friedmann-AS-2}
& H^2 = \frac{1}{3M_\mathrm{P}^2}\left\{\frac{\dot{\phi}^2}{2}+V(\phi)+\frac{H^4}{2}\left[e_0\left(\tfrac{\beta \dot{\phi}}{2HM_{\mathrm{P}}}\right)+b_0\left(\tfrac{\beta \dot{\phi}}{2HM_{\mathrm{P}}}\right)\right]\right\} \,.
\end{align}
Note that for a generic inflationary model, the solution of Eqs.~\eqref{KGF-AS-2}--\eqref{Friedmann-AS-2} is not a solution of the full system of equations (including Maxwell's equations for the gauge field), since in the full system the assumptions underlying Eqs.~\eqref{KGF-AS-2}--\eqref{Friedmann-AS-2}, namely $H=\text{const}$ and $\xi=\text{const}$, are not necessarily satisfied. Therefore, we will often refer to this solution as the ``enforced'' AS (EAS) solution, which is characterized by the fact that we insist on (or ``enforce'') the specific functional dependence encoded in the functions $e_0$, $g_0$, and $b_0$. We study the stability of the EAS solution in a specific inflationary model in section~\ref{sec:realistic}. On the other hand, in the particular case of constant background quantities, considered in detail in Sec.~\ref{sec:const_back}, the AS solution is an exact solution of the full system of equations, which can be realized by choosing the right initial conditions.%
\footnote{Strictly speaking, the conditions of constant $H$ and $\dot\phi$ cannot be simultaneously satisfied, as a rolling inflaton will induce a time dependence in the Hubble parameter. This approximation becomes exact in the limit in which $\beta\to \infty$ (\textit{i.e.}, $f \to 0$) and $\dot{\phi} \to 0$, while  $H$ and $\xi$ are finite and constant.}
Nonetheless, a main result of our analysis will be that, even though the AS solution is an exact solution of Eq.~(\ref{KGF-AS-2}) for constant $H$ and $\xi$, it turns out to be unstable against arbitrarily small perturbations and therefore only has a finite survival time, as we will demonstrate in detail.
In Table~\ref{tab:notations}, we summarize our notations for the different types of solutions considered in the next sections as well as the methods used to obtain them.

\begin{table}[t!]
\renewcommand{\arraystretch}{1.5}
\begin{tabular}{|l|l|l|}
\hline
\multicolumn{3}{|l|}{Sec.~\ref{sec:const_back}: Constant gradient $V'$, constant Hubble rate $H$, Friedmann equation ignored} \\\hline
Solution      & $\xi$ parameter                                      & Methods \\\hline
AS            & $\bar{\xi}\left(t\right)$                            & $\bar{\xi}\left(t\right) = \xi_0 = \textrm{const}$ is an exact solution \\
Full solution & $\bar{\xi}\left(t\right) + \delta\sm\xi\left(t\right)$ & \textbullet~Linear regime at early times:\\[-1.5ex]
              &                                                      & Linear perturbation theory and linearized GEF\\[-1ex]
              &                                                      & \textbullet~Nonlinear regime at late times:\\[-1.5ex]
              &                                                      & Full GEF and mode-by-mode solution with GEF input\\\hline
              \hline
\multicolumn{3}{|l|}{Sec.~\ref{sec:realistic}: Constant gradient $V'$, dynamical Hubble rate $H$, Friedmann equation included} \\\hline
Solution      & $\xi$ parameter                                      & Methods \\\hline
Enforced AS   & $\bar{\xi}\left(t\right)$                            & Numerical solution of Eqs.~\eqref{KGF-AS-2}--\eqref{Friedmann-AS-2} \\
Full solution & $\xi\left(t\right)$                                  & Full GEF and mode-by-mode solution with GEF input \\\hline
\end{tabular}
\caption{Overview of the solutions studied in Sec.~\ref{sec:const_back} and  \ref{sec:realistic} and the methods used to obtain them.
\label{tab:notations}}
\end{table}


\subsection{Parameter choice}
\label{subsec:parameters}

Let us now specify the model parameter values that we are going to be interested in, \textit{i.e.}, the region in parameter space corresponding to the strong-backreaction regime. To do so, we assume that the system is initialized in phase space either in the AS solution or at least sufficiently close to it, so that we can employ the equations of motion~\eqref{KGF-AS-2}--\eqref{Friedmann-AS-2}.

Specifically, we shall impose two conditions. On the one hand, the backreaction in the Klein--Gordon equation must be strong, meaning that the additional friction from the gauge field dominates over the Hubble friction term and hence governs the evolution of the inflaton field. On the other hand, the contribution of the produced gauge fields to the total energy density of the Universe needs to remain small compared to that of the inflaton, so that we can still realize a stage of accelerated (inflationary) expansion with the effective equation of state parameter $w=p/\rho<-1/3$. In order to give a quantitative meaning to these two conditions, we introduce two parameters, $\delta_{\mathrm{KG}}$ and $\delta_{\mathrm{F}}$, which measure the strength of the backreaction in the Klein--Gordon and Friedmann equations, respectively,
\begin{align}
\delta_{\mathrm{KG}} & = \frac{|(\beta/M_{\mathrm{P}})\langle \bm{E}\cdot\bm{B}\rangle |}{|3H\dot{\phi}|}=\frac{1}{6}\left(\frac{\beta H}{M_{\mathrm{P}}}\right)^{2}\left|\frac{g_0(\xi)}{\xi}\right| \,, \\
\delta_{\mathrm{F}} & =
\,\frac{\frac{1}{2}(\langle \bm{E}^2\rangle+\langle \bm{B}^2\rangle)}{3H^2 M_{\mathrm{P}}^2}\,=\frac{1}{6}\left(\frac{H}{M_{\mathrm{P}}}\right)^{2}\left[e_0(\xi)+b_0(\xi)\right] \,.
\end{align}
Then, the conditions determining the desired parameter range can be formulated as
\begin{equation}
\label{conditions-AS-exists}
    \delta_{\mathrm{KG}}\gg 1 \,, \qquad \delta_{\mathrm{F}}\ll 1 \,.
\end{equation}
These two conditions are independent and satisfied across an extended volume in the three-dimensional parameter space spanned by $\beta$, $\xi$, and $H$. Still, it will be helpful to define a benchmark in the sense of an ``optimal parameter choice'' determined by the condition
\begin{equation}
\label{optimal-choice}
\sqrt{\delta_{\mathrm{KG}}\,\delta_{\mathrm{F}}} = 1  \,.
\end{equation}
This choice gives the central section of the relevant parameter range where the backreaction is strong in the Klein--Gordon equation and, at the same time, small in the Friedmann equation. We expect that even away from the optimal parameter choice, as long as the conditions \eqref{conditions-AS-exists} are satisfied, we will obtain qualitatively the same results.

The condition \eqref{optimal-choice} allows to eliminate one of the parameters, \textit{e.g.}, the Hubble rate, in terms of the two other parameters,
\begin{equation}
\frac{H}{M_{\mathrm{P}}} = \frac{1}{\beta^{1/2}} \left(\frac{36\,\xi}{g_0(\xi)[e_0(\xi)+b_0(\xi)]}\right)^{1/4}.
\end{equation}
To obtain an intuition for the analytical dependence on the parameter $\xi$, we use the approximate expressions in Eq.~\eqref{e0b0g0-approx}, which allow us to write the more explicit expression
\begin{equation}
\frac{H}{M_{\mathrm{P}}}=\frac{2^{33/4}\pi}{1575^{1/4}} \frac{\xi^2 e^{-\pi\xi}}{\sqrt{\beta}}\left(1+\frac{7}{8\xi^2}\right)^{-1/4} \,.
\end{equation}
Moreover, in the range of $\xi$ values $5\lesssim \xi\lesssim 10$, which is the most interesting for the present study, we find a simple empirical relation, \textit{i.e.}, a fit formula,
\begin{equation}
\label{H-optimal}
\frac{H}{M_{\mathrm{P}}} = 2 \times 10^{-7}\, \left(\frac{100}{\beta}\right)^{1/2} \exp[-2.85(\xi-7)],
\end{equation}
which reproduces the exact result up to an error of a few percent. This is good enough for us; the condition in Eq.~\eqref{optimal-choice} is not an exact requirement, anyway. It merely serves the purpose of providing us with guidance as to where in parameter space we can expect the strong-backreaction regime of axion inflation to be realized. Unless specified otherwise, we will therefore use the relation \eqref{H-optimal} in all computations in the remainder of this work.


\section{Constant background quantities}
\label{sec:const_back}

Let us start our discussion of the AS solution considering the simple case of constant Hubble rate and constant inflaton potential gradient, $H = \textrm{const}$ and $V'\left(\phi\right) = \text{const}$. In this case, the Universe expands exponentially (de Sitter spacetime), and we effectively disregard the Friedmann equation.\footnote{As we already remarked, doing so, the Friedmann equation is violated by the fact that the total energy density cannot be perfectly constant for $V'\left(\phi\right) \neq 0$; in this section we assume a regime in which this violation is sufficiently small so not to substantially impact our results. This equation is fully considered in the evolution studied in the next section.} At the same time, the Klein--Gordon equation admits a solution with constant inflaton velocity. Indeed, setting $\ddot{\phi}=0$ in Eq.~\eqref{KGF-AS-2} and expressing everything in terms of the $\xi$ parameter, which is also constant, we obtain the following equation:
\begin{equation}
\label{AS-const}
\frac{6H^2 M_\mathrm{P}}{\beta}\,\xi + V' = -\frac{\beta}{M_{\mathrm{P}}}H^4 \, g_0 \left(\xi\right) \,.
\end{equation}
For given values of the axion--vector coupling $\beta$, Hubble rate $H$, and potential gradient $V'$, this equation can be solved for the associated constant value of $\xi$. Thus, in this simple case, the AS solution turns out to be $\xi\left(t\right) = \xi_0 = \text{const}$; see Table~\ref{tab:notations}. It is important to note that this is not only the solution of the approximate Eq.~\eqref{KGF-AS-2}, but also a particular solution of the full system of equations, including the EOM for the gauge-field modes. Since $\xi = \text{const}$, the gauge-field mode functions have the form \eqref{A-Whittaker} and, therefore, the simple relations in Eq.~\eqref{e0b0g0} are exact. In particular, this means that a system prepared \textit{exactly} in this state will remain in it forever. It is interesting, however, to study the stability of this solution and consider the evolution of small perturbations around it. In what follows, we shall denote all quantities in the AS solution by a bar, \textit{e.g.}, $\bar{\xi}$.


\subsection{Linear perturbation theory}
\label{subsec:linearpt}

In order to study the stability of the AS solution, let us construct a linear perturbation theory for deviations from this solution following the same strategy as in Ref.~\cite{Peloso:2022ovc}. We write the perturbed quantities as
\begin{equation}
\phi\left(\tau\right)=\bar{\phi}\left(\tau\right)+\delta\sm\phi(\tau) \,, \qquad A_\lambda\left(\tau,k\right)=\bar{A}_\lambda\left(\tau,k\right)+\delta\sm A_\lambda\left(\tau,k\right) \,.
\end{equation}
The background quantities evolve according to equations similar to Eqs.~\eqref{A_2} and \eqref{KGF-2},
\begin{align}
\label{A_2-AS}
& \frac{\partial^2 \bar{A}_\lambda(\tau, k)}{\partial\tau^2}+\left[k^2-\lambda k\frac{\beta}{M_{\mathrm{P}}}\frac{d\bar{\phi}}{d\tau}\right]\bar{A}_\lambda(\tau, k) = 0 \,, \\
\label{KGF-sp-AS}
& \frac{d^2\bar{\phi}}{d\tau^2}+2aH\frac{d\bar{\phi}}{d\tau}+a^2 V'(\bar{\phi}) = -\frac{\beta}{4\pi^2 a^2 M_\mathrm{P}}\int\limits_0^{k_{\mathrm{h}}}dk\,k^3 \frac{\partial}{\partial \tau}\sum\limits_{\lambda=\pm}\lambda\big|\bar{A}_{\lambda}(\tau,k)\big|^2.
\end{align}
Subtracting Eqs.~\eqref{A_2-AS}, \eqref{KGF-sp-AS} from Eqs.~\eqref{A_2}, \eqref{KGF-2}, respectively, and keeping only perturbation terms up to linear order, we obtain the system of EOMs for the perturbations,
\begin{align}
\label{A-pert}
& \frac{\partial^2 \delta\sm A_\lambda(\tau,k)}{\partial\tau^2}+\left[k^2-\lambda k\frac{\beta}{M_{\mathrm{P}}}\frac{d\bar{\phi}}{d\tau}\right]\delta\sm A_\lambda(\tau, k)=\lambda k\frac{\beta}{M_{\mathrm{P}}}\bar{A}_\lambda(\tau,k)\frac{d\delta\sm\phi}{d\tau} \,, \\
\label{KGF-pert}
& \frac{d^2\delta\sm\phi}{d\tau^2}+2aH\frac{d\delta\sm\phi}{d\tau}+a^2 V''(\bar{\phi})\delta\sm\phi=-\frac{\beta}{2\pi^2 a^2 M_\mathrm{P}}\int\limits_0^{k_{\mathrm{h}}}dk\,k^3 \frac{\partial}{\partial \tau}\sum\limits_{\lambda=\pm}\lambda \operatorname{Re}[\bar{A}_{\lambda}(\tau,k)\delta\sm A_{\lambda}^{\ast}(\tau,k)] \,.
\end{align}

The solution of Eq.~\eqref{A-pert} can be formally expressed as,
\begin{equation}
    \label{A-sol-Green}
    \delta\sm A_{\lambda}(\tau,k)=\lambda k\frac{\beta}{M_{\mathrm{P}}}\int\limits_{-\infty}^{\tau}d\tau'\,G_{\lambda,k}(\tau,\tau')\bar{A}_\lambda(\tau',k)\frac{d\delta\sm\phi(\tau')}{d\tau'} \,,
\end{equation}
where the Green function $G_{\lambda,k}(\tau,\tau')$ is a solution of the differential equation
\begin{equation}
    \left[\frac{\partial^2}{\partial \tau^2} +k^2-\lambda k\frac{\beta}{M_{\mathrm{P}}}\frac{d\bar{\phi}}{d\tau} \right] G_{\lambda,k}(\tau,\tau')=\delta(\tau-\tau') \,.
\end{equation}
The differential operator acting on the Green function is exactly the same as the differential operator in Eq.~\eqref{A_2-AS}. $\bar{A}(\tau,k)$ is hence the solution of the corresponding homogeneous equation. A second linearly independent solution is $\bar{A}^{\ast}(\tau,k)$, since the mode equation has real coefficients. This allows us to construct the retarded Green function as follows:
\begin{equation}
    \label{Green-function}
     G_{\lambda,k}(\tau,\tau')=i\left[\bar{A}_\lambda(\tau,k)\bar{A}_\lambda^{\ast}(\tau',k)-\bar{A}_\lambda^{\ast}(\tau,k)\bar{A}_\lambda(\tau',k)\right]\theta(\tau-\tau') \,,
\end{equation}
where we used the fact that the mode functions are normalized in such way that their Wronskian equals
\begin{equation}
    \label{Wronskian}
    \frac{\partial\bar{A}_\lambda(\tau,k)}{\partial \tau}\bar{A}_\lambda^{\ast}(\tau,k)-\frac{\partial\bar{A}_\lambda^{\ast}(\tau,k)}{\partial \tau}\bar{A}_\lambda(\tau,k)=-i.
\end{equation}

Substituting Eq.~\eqref{A-sol-Green} into Eq.~\eqref{KGF-pert}, we get the source term on the right-hand side,
\begin{equation}
    \text{RHS}=\frac{1}{2\pi^2}\left(\frac{\beta}{a M_\mathrm{P}}\right)^2\int\limits_0^{k_{\mathrm{h}}} dk\, k^4 \frac{\partial}{\partial\tau}\int\limits_{-\infty}^{\tau} d\tau'\,\frac{d\delta\sm\phi(\tau')}{d\tau'}\sum\limits_{\lambda=\pm}\operatorname{Im}\left[\bar{A}_\lambda^2(\tau,k)\bar{A}^{\ast 2}_\lambda(\tau',k)\right] \,.
\end{equation}
Taking into account that $\operatorname{Im}[\ldots] = 0$ when $\tau=\tau'$ in this expression, we evaluate the derivative with respect to $\tau$ and finally obtain the EOM for the scalar-field perturbation,
\begin{multline}
\label{KGF-pert-2}
\frac{d^2\delta\sm\phi}{d\tau^2}+2aH\frac{d\delta\sm\phi}{d\tau}+a^2 V''(\bar{\phi})\delta\sm\phi \\
= \left(\frac{\beta}{\pi a M_\mathrm{P}}\right)^2\int\limits_0^{k_{\mathrm{h}}} dk\, k^4 \int\limits_{-\infty}^{\tau} d\tau'\,\frac{d\delta\sm\phi(\tau')}{d\tau'}\sum\limits_{\lambda=\pm}\operatorname{Im}\left[\bar{A}_\lambda(\tau,k)\frac{\partial \bar{A}_\lambda(\tau,k)}{\partial \tau}\bar{A}^{\ast\,2}_\lambda(\tau',k)\right] \,.
\end{multline}

In Eq.~\eqref{KGF-pert-2}, we kept the most general form of the mode functions. Next, in order to simplify Eq.~\eqref{KGF-pert-2}, we assume that the gauge-field mode functions can be represented as
\begin{equation}
\label{barA-ansatz}
\bar{A}_\lambda\left(\tau, k\right) = \frac{1}{\sqrt{2k}}\,\mathcal{W}_\lambda \left(x\right) \,,\qquad x = -k\tau=\frac{k}{aH} \,,
\end{equation}
where $\mathcal{W}_\lambda \left(x\right)$ is an arbitrary function of $x$ and $\lambda$ for the time being.
Both the exact solution in Eq.~\eqref{A-Whittaker} and the approximate one in Eq.~\eqref{A-plus-approx} can be represented in this way. Moreover, following Ref.~\cite{Peloso:2022ovc}, we look for power-law solutions of the EOM \eqref{KGF-pert-2}. We therefore choose the following ansatz for the scalar-field perturbation:
\begin{equation}
\label{dphi-ansatz}
\delta\sm\phi\left(\tau\right) = C \left(-\tau\right)^{-\zeta} \,,
\end{equation}
where $C$ and $\zeta$ are constant. Once this ansatz is inserted in Eq.~(\ref{KGF-pert-2}), and the allowed $\zeta$ are found, we exploit the fact that, for any solution $\zeta$, also its complex conjugate $\zeta^*$ is a solution, to obtain a real inflaton perturbation, 
\begin{equation}
\label{growing-oscillations}
\delta\sm\phi\left(\tau\right) = C  \left[ \left(-\tau\right)^{-\zeta} + \left(-\tau\right)^{-\zeta^*} \right]  = C  \left[ H^\zeta e^{\zeta N} + H^{\zeta^*} e^{\zeta^* N}\right] = \tilde{C}\,e^{\operatorname{Re}\left(\zeta\right) N} \cos\left[\operatorname{Im}\left(\zeta\right) N+\varphi_0\right] \,.
\end{equation}
where we have introduced the number of $e$-folds, $N \equiv \ln a$. 
Writing $\delta\sm\phi$ in this way, we can identify $\operatorname{Re}(\zeta)$ with the growth rate and $\operatorname{Im}(\zeta)$ with the angular oscillation frequency of $\delta\sm\phi$ as a function of the number of $e$-folds. In fact, the most general linearized perturbation will be a linear combination of modes of the form (\ref{growing-oscillations}), where each term has one allowed $\zeta_n$ (with its complex conjugate) and the coefficients $C_n$ are obtained from the initial condition for the perturbation.

Now, we set $V''=0$ (since in our case $V'=\text{const}$), substitute Eqs.~\eqref{barA-ansatz} and \eqref{dphi-ansatz} into Eq.~\eqref{KGF-pert-2}, and perform the change of integration variables $\tau' \to x'=-k\tau'$ and $k\to x=-k\tau$. After that, the combination $C\left(-\tau\right)^{-\zeta-2}$ appears in all terms on the left- and right-hand sides. Canceling this $\tau$-dependence, we obtain the following equation for $\zeta$:
\begin{equation}
\label{eq-zeta-general}
\zeta\left(\zeta+3\right) = -\frac{\zeta}{4\pi^2}\left(\frac{\beta H}{M_\mathrm{P}}\right)^2 \int\limits_0^{2\bar{\xi}} dx\, x^{3+\zeta} \int\limits_{x}^{+\infty} \frac{dx'}{\left(x^{\prime}\right)^{\zeta+1}} \sum\limits_{\lambda=\pm}\operatorname{Im}\left[\mathcal{W}_\lambda(x)\mathcal{W}'_\lambda(x)\mathcal{W}^{\ast\,2}_\lambda(x')\right] \,,
\end{equation}
where $\mathcal{W}'_\lambda(x)=d\mathcal{W}_\lambda(x)/dx$. Up to now, we were able to work with a general function $\mathcal{W}_\lambda(x)$. However, in order to determine the allowed power-law exponents $\zeta_n$, one needs to  specify the function $\mathcal{W}_\lambda(x)$ and solve Eq.~\eqref{eq-zeta-general} numerically. In the case of constant $H$ and $\bar{\xi}$, which we consider here, the function $\mathcal{W}_\lambda(x)$ can be extracted from Eq.~\eqref{A-Whittaker},
\begin{equation}
\mathcal{W}_\lambda(x) = e^{\pi\lambda\bar{\xi}/2}\, W_{-i\lambda\bar{\xi},1/2}\left(-2ix\right) \,.
\end{equation}
This equation may be simplified following Ref.~\cite{Peloso:2022ovc}. Firstly, we take into account only the enhanced $\lambda=+$ gauge polarization. Secondly, we replace the upper integration limit in the second integral by $2\bar{\xi}$, which is motivated by the fact that for $x'>2\bar{\xi}$ the integrand is no longer enhanced, but it actually becomes a rapidly oscillating function that integrates to a negligible amount. Thirdly, if we use the approximate form of the mode function $A_{+}$ in Eq.~\eqref{A-plus-approx}, we obtain an approximate equation for $\zeta$ of the form 
\begin{equation}
\label{eq-zeta-approx}
R\left(\zeta\right) \equiv \left(\frac{\beta H}{M_\mathrm{P}}\right)^2 \frac{315}{2^{15}\pi^2}\frac{e^{2\pi\bar{\xi}}}{\bar{\xi}^5}\frac{1}{\left(2\zeta-1\right)\left(2\zeta+7\right)}\left[\frac{\Gamma\left(2\zeta+8\right)}{\Gamma\left(9\right)\left(8\bar{\xi}\right)^{2\zeta-1}}-1\right] = 1 \,,
\end{equation}
which agrees with Eq.~(3.14) in Ref.~\cite{Peloso:2022ovc}, expressed in the notation of the present paper.


\subsection{Linearized gradient expansion formalism}

The gradient expansion formalism (GEF) introduced in Sec.~\ref{subsec:GEF} allows us to find an exact numerical solution for the system of coupled inflaton and gauge-field EOMs in the strong-backreaction regime. For constant background quantities $H$ and $V'$, the system of equations can be further simplified. First of all, since the inflaton field $\phi$ itself does not appear in the system and only its derivative $\dot{\phi}$ is involved, it is convenient to use the parameter $\xi$ in Eq.~\eqref{xi} as a new field variable. This renders the Klein--Gordon equation a first-order ordinary differential equation, like all the equations of the GEF. Moreover, since $H=\text{const}$, it is more convenient to work with the number of $e$-folds, $N=\ln a=Ht$ as a time variable, instead of physical time $t$. This leads us to:
\begin{equation}
\label{KGF-xi}
\frac{d\xi}{dN} + 3\,\xi - v = -\frac{b^2}{2} G^{(0)} \,, \qquad v \equiv -\frac{\beta V'}{2H^2M_\mathrm{P}} \,, \qquad b \equiv \frac{\beta H}{M_\mathrm{P}} \,,
\end{equation}
where $v$ and $b$ are dimensionless parameters accounting for the dragging force caused by the potential gradient and the axion--vector coupling, respectively. This equation needs to be supplemented by the tower of equations that govern the gauge-field evolution:
\begin{align}
\label{E_n_const}
& \frac{dE^{(n)}}{dN}+(n+4)E^{(n)}-4\xi G^{(n)}+2G^{(n+1)} = \frac{(2\xi)^{n+4}}{4\pi^2}\frac{d\ln \kappa_\mathrm{h}}{dN}\sum\limits_{\lambda=\pm}\lambda^n E_\lambda \,, \\
\label{G_n_const}
& \frac{dG^{(n)}}{dN}+(n+4)G^{(n)}-2\xi B^{(n)}+B^{(n+1)}-E^{(n+1)} = \frac{(2\xi)^{n+4}}{4\pi^2}\frac{d\ln \kappa_\mathrm{h}}{dN}\sum\limits_{\lambda=\pm}\lambda^{n+1} G_\lambda \,, \\
\label{B_n_const}
& \frac{dB^{(n)}}{dN}+(n+4)B^{(n)}-2G^{(n+1)} = \frac{(2\xi)^{n+4}}{4\pi^2}\frac{d\ln \kappa_\mathrm{h}}{dN}\sum\limits_{\lambda=\pm}\lambda^n B_\lambda \,,
\end{align}
where $E_\lambda$, $G_\lambda$, and $B_\lambda$ are functions of the production parameter $\xi$ given in Eqs.~\eqref{E_lambda}--\eqref{B_lambda}. In Eqs.~\eqref{E_n_const}--\eqref{B_n_const}, moreover, we introduced the dimensionless bilinear functions
\begin{equation}
    \label{dimless-bilinear}
    E^{(n)}=\frac{\mathcal{E}^{(n)}}{H^{n+4}} \,, \qquad G^{(n)}=\frac{\mathcal{G}^{(n)}}{H^{n+4}} \,, \qquad
    B^{(n)}=\frac{\mathcal{B}^{(n)}}{H^{n+4}} \,,
\end{equation}
the dimensionless momentum of the horizon-crossing mode
\begin{equation}
\label{kappa-h}
\kappa_\mathrm{h}\left(N\right) = \frac{k_\mathrm{h}}{H}=\max\limits_{N'<N}\left[2\xi\left(N'\right)e^{N'}\right] \,,
\end{equation}
and its derivative
\begin{equation}
\label{dkappa_h-dN}
\frac{d\ln \kappa_\mathrm{h}}{dN}=\left(\frac{1}{\xi}\frac{d\xi}{dN}+1\right) \theta\left(\frac{1}{\xi}\frac{d\xi}{dN}+1\right) \theta\left(2\xi\,e^N - \kappa_\mathrm{h}\left(N\right)\right) \,,
\end{equation}
where $\theta(x)$ is the Heaviside unit step function.

Although this system can be directly employed to study the true solution in the strong-backreaction regime, it is also instructive to linearize it for small deviations from the AS solution. Denoting all quantities in the AS solution by bars and small deviations by $\delta$'s, we obtain the following system of equations, which define what we shall refer to as the linearized gradient expansion formalism (LGEF):
\begin{align}
\label{KGF-xi-delta}
& \frac{d\delta\sm\xi}{dN}+3\delta\sm\xi =-\frac{b^2}{2} \delta\sm G^{(0)} \,, \\
\label{E_n_const-delta}
& \frac{d\delta\sm E^{(n)}}{dN}+(n+4)\delta\sm E^{(n)}-4\bar{\xi}\, \delta\sm G^{(n)}-4\bar{G}^{(n)}\,\delta\sm\xi+2\delta\sm G^{(n+1)}= \delta\sm S_E \,, \\
\label{G_n_const-delta}
& \frac{d\delta\sm G^{(n)}}{dN}+(n+4)\delta\sm G^{(n)}-2\bar{\xi} \delta\sm B^{(n)}-2\bar{B}^{(n)}\,\delta\sm \xi+\delta\sm B^{(n+1)}-\delta\sm E^{(n+1)} = \delta\sm S_G \,, \\
\label{B_n_const-delta}
& \frac{d\delta\sm B^{(n)}}{dN}+(n+4)\delta\sm B^{(n)}-2\delta\sm G^{(n+1)} = \delta\sm S_B \,, 
\end{align}
where
\begin{align}
\delta\sm S_E & = \frac{(2\bar{\xi})^{n+4}}{4\pi^2}\sum\limits_{\lambda=\pm}\lambda^n \bigg\{\frac{E_\lambda(\bar{\xi})}{\bar{\xi}}\left[(n+4)\delta\sm \xi+\frac{d\delta\sm\xi}{dN}\right]+\frac{dE_\lambda(\bar{\xi})}{d\bar{\xi}}\delta\sm \xi\bigg\} \,, \\
\delta\sm S_G & = \frac{(2\bar{\xi})^{n+4}}{4\pi^2}\sum\limits_{\lambda=\pm}\lambda^{n+1} \bigg\{\frac{G_\lambda(\bar{\xi})}{\bar{\xi}}\left[(n+4)\delta\sm \xi+\frac{d\delta\sm\xi}{dN}\right]+\frac{dG_\lambda(\bar{\xi})}{d\bar{\xi}}\delta\sm \xi\bigg\} \,, \\
\delta\sm S_B & = \frac{(2\bar{\xi})^{n+4}}{4\pi^2}\sum\limits_{\lambda=\pm}\lambda^n \bigg\{\frac{B_\lambda(\bar{\xi})}{\bar{\xi}}\left[(n+4)\delta\sm \xi+\frac{d\delta\sm\xi}{dN}\right]+\frac{dB_\lambda(\bar{\xi})}{d\bar{\xi}}\delta\sm \xi\bigg\} \,. 
\end{align}
This system is also infinite in principle, and, in order to use it in practice, one has to truncate it at some order $n_\mathrm{cut}$. The simplest way to do so is to assume that, for all orders larger than $n_\mathrm{cut}$, the bilinear functions exactly coincide with the background values in the AS solution, \textit{i.e.},
\begin{equation}
\label{LGEF-truncation}
\delta\sm E^{(n)}=\delta\sm G^{(n)}=\delta\sm B^{(n)}=0,\qquad n>n_\mathrm{cut}.
\end{equation}
The advantage of the LGEF compared to other methods is that it leads to a system of linear ODEs with constant coefficients. Its solution can easily be found by methods of linear algebra. In particular, the ansatz $\delta\sm\xi\propto e^{\zeta N}$ (the same as in the previous subsection), and similarly for all perturbations of bilinear functions, reformulates the problem from studying the evolution in time to just finding eigenvalues of the matrix of the linear system, \textit{i.e.}, to a purely algebraic task. In practice, this turns out to be the simplest approach.


\subsection{Results and discussion}

We shall now discuss our numerical results obtained for the case of constant Hubble rate $H$ and constant potential gradient $V'$. As noted above, the AS solution is a true solution of the Maxwell and Klein--Gordon equations. This means that, if we perfectly fine-tune the initial conditions for $\xi$ and all gauge-field bilinear quantities to be in the AS solution, the system will remain in this solution indefinitely. Perfectly fine-tuned initial conditions are, however, of little interest. In fact, they are even impossible to achieve in any numerical study with finite numerical precision. 
In what follows, we will therefore slightly detune the initial conditions and study the dynamical evolution of the system away from the AS solution. As we will find that the AS solution is unstable, we sometimes denote this as the ``decay of the AS solution'', which also features prominently in the title of this paper.

In this subsection, we apply all three approaches discussed above in order to study the stability of the AS solution with respect to small perturbations. The linear perturbation theory and the LGEF allow us to determine the spectrum of Lyapunov exponents $\zeta$ and, thus, to capture all possible scenarios for the evolution of the system at once, however, only in the regime of small perturbations. On the other hand, the full GEF allows us to get an exact numerical solution of the system for some specified initial conditions that is valid also for large deviations from the AS solution. Therefore, these methods are complementary to each other and allow us to study the same system from different angles.

\begin{figure}
\centering
\includegraphics[width=0.73\linewidth]{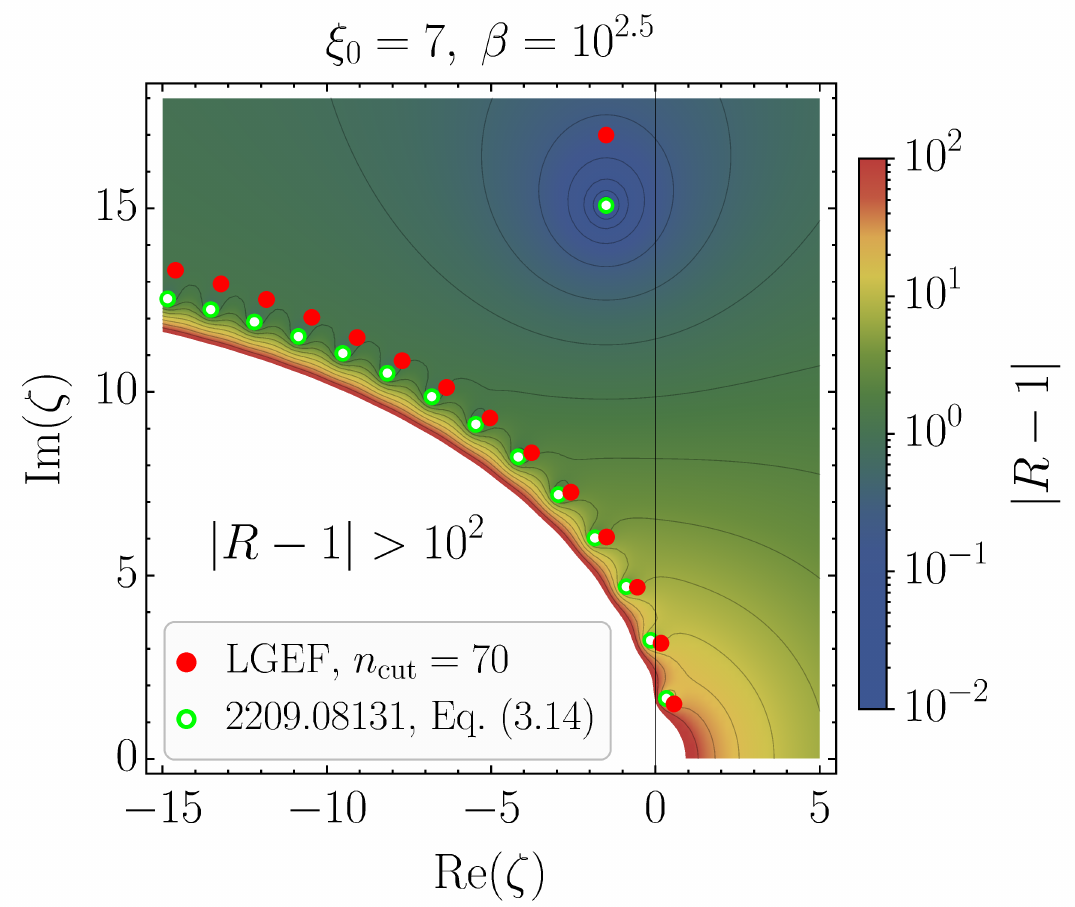}
\caption{Eigenvalues of the LGEF system with $n_\mathrm{cut}=70$ (red dots) and solutions of Eq.~\eqref{eq-zeta-approx}, which agrees with Eq.~(3.14) in Ref.~\cite{Peloso:2022ovc} (green circles), in the complex $\zeta$ plane for $\xi_0=7$ and $\beta=10^{2.5}$. The contour plot in the background shows the absolute deviation from equality in Eq.~\eqref{eq-zeta-approx}. In the white region, $\left|R-1\right| > 10^{2}$, exceeding the scale of the color code. For each solution $\zeta$, also the complex conjugate $\zeta^*$ is a solution. In this figure, only the solutions with positive imaginary parts are shown. \label{fig-complex-plane}}
\end{figure}

\subsubsection{Lyapunov exponents}

It is instructive to first work out the spectrum of Lyapunov exponents $\zeta$ for our physical system. For definiteness, we perform the numerical analysis in the region of parameter space  $\xi_0\in[5,\,9]$, $\beta\in[10^{1.5},\,10^{3.5}]$, and $H$ determined by Eq.~\eqref{H-optimal}. For the benchmark point with $\beta=10^{2.5}$ and $\xi_0=7$ (which is the central point of the specified parameter range), this spectrum is shown in Fig.~\ref{fig-complex-plane} in the form of a sequence of red dots in the complex plane for $\zeta$. These points have been found by using the LGEF truncated at $n_{\mathrm{cut}}=70$. We want to compare the results of this numerical integration with the solutions of Eq.~\eqref{eq-zeta-general} for $\zeta$. This equation, which follows from the linear perturbation theory, contains integrals of highly oscillatory special functions, and it turns out that finding its solutions is computationally very costly. We have checked that, for the root with the greatest real part, $\zeta_1$, the numerical solution of Eq.~\eqref{eq-zeta-general} is in perfect agreement with the LGEF result presented in the figure. This root is of great importance since, having the greatest real part, it is the one that controls the growth rate of the instability at late times; see Eq.~(\ref{dphi-ansatz}). A much quicker comparison of the eigenvalues obtained from the LGEF system with the analytical computation can be done in terms of the approximate equation~\eqref{eq-zeta-approx} (\textit{i.e.}, Eq.~(3.14) in Ref.~\cite{Peloso:2022ovc}), which is easier to solve. We show the roots of this equation by green empty circles in Fig.~\ref{fig-complex-plane}. The background color of this figure is the density plot of the absolute value of $R-1$, the difference between the function on the left-hand side of Eq.~\eqref{eq-zeta-approx} and unity. As evident from Fig.~\ref{fig-complex-plane}, the approximate equation~\eqref{eq-zeta-approx} allows us to obtain the spectrum of the Lyapunov exponents $\zeta$ with good accuracy. In particular, it also shows that among the $\zeta$ values there is at least one with positive real part meaning that the AS solution is unstable; this confirms the findings of Ref.~\cite{Peloso:2022ovc}. Fig.~\ref{fig-complex-plane} also shows that there is a one-to-one correspondence between the solutions obtained from the LGEF method and the solutions that we were able to derive in the context of linear perturbation theory and starting from the ansatz in Eq.~\eqref{dphi-ansatz}. This observation serves as another (\textit{a posteriori}) justification for the ansatz in Eq.~\eqref{dphi-ansatz} and confirms that we did not overlook any solutions in our linear-perturbation-theory analysis in Sec.~\ref{subsec:linearpt}.

\begin{figure}
\centering
\includegraphics[width=0.48\linewidth]{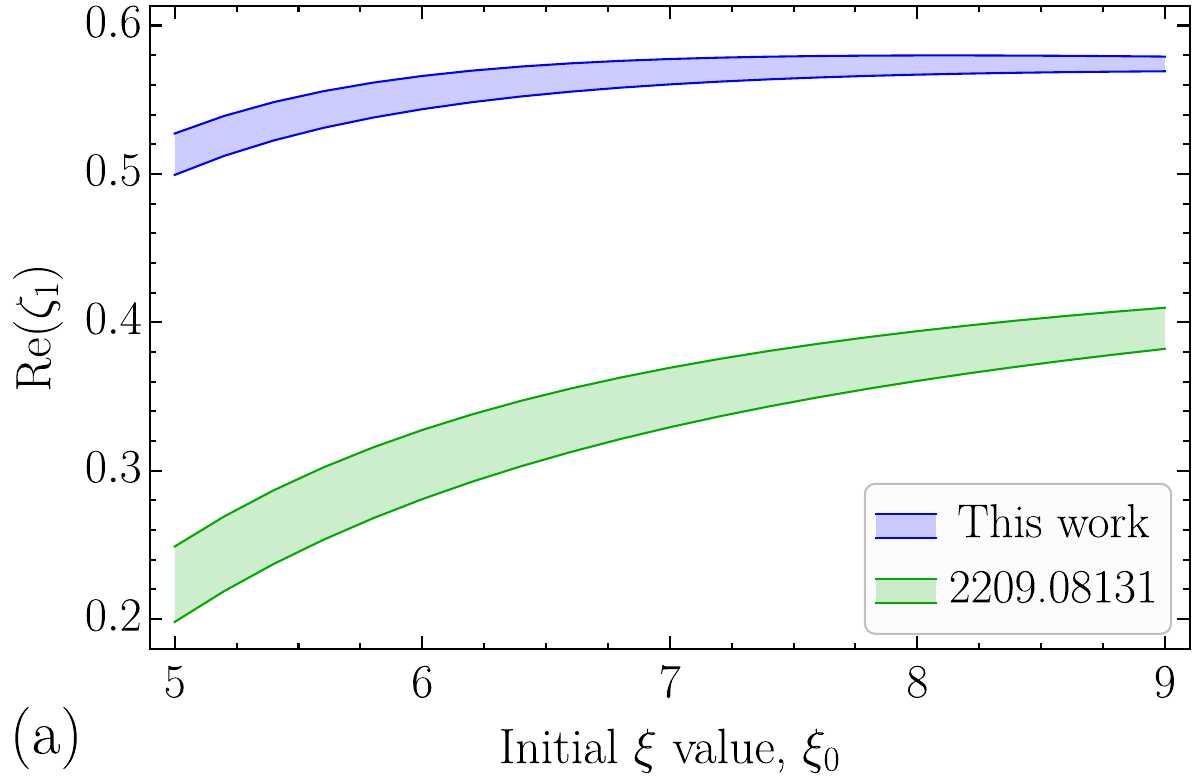}
\hspace{3mm}
\includegraphics[width=0.48\linewidth]{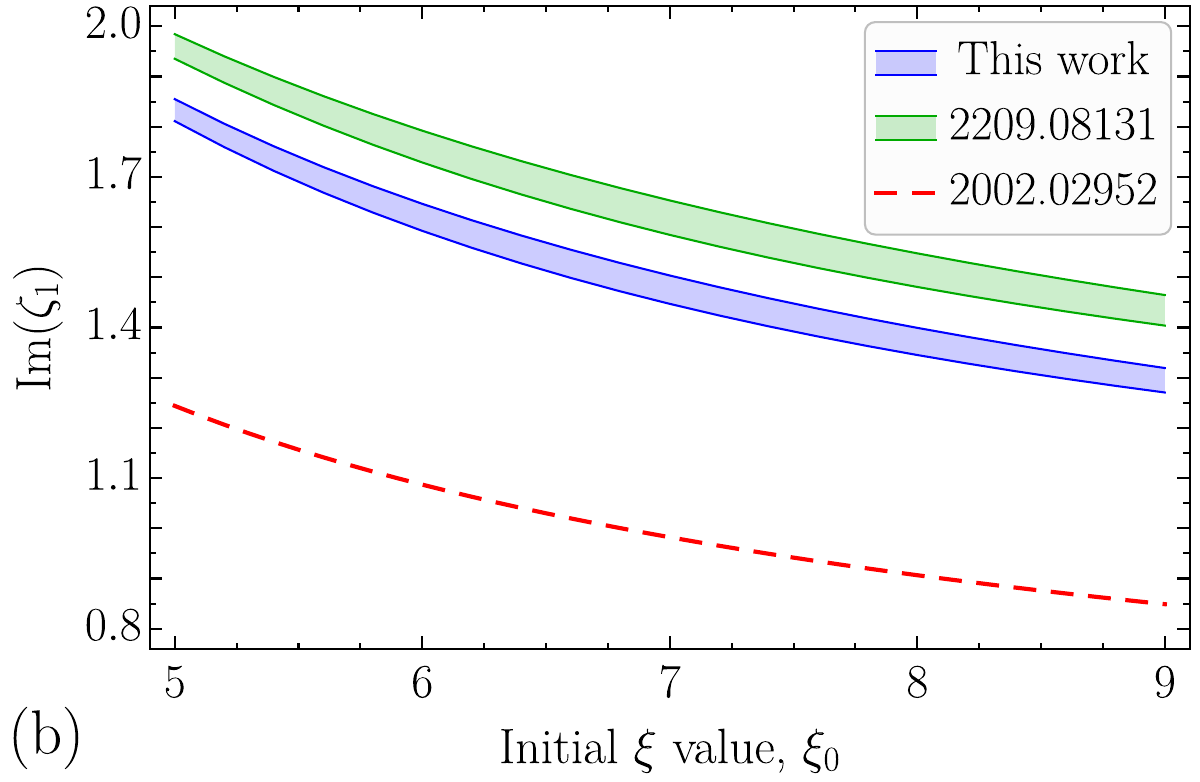}
\caption{(a)~Real and (b)~imaginary parts of the eigenvalue $\zeta_{1}$, corresponding to the fastest-growing mode, as functions of $\xi_0$. The bands of finite width reflect different values of $\beta\in[10^{1.5},\,10^{3.5}]$. The blue band follows from the LGEF, while the green band shows the solution of Eq.~\eqref{eq-zeta-approx}. The red dashed line in panel~(b) is the analytical estimate in Ref.~\cite{Domcke:2020zez}; see Eq.~\eqref{estimate-Domcke}.
}
\label{fig-zeta-on-xi}
\end{figure}

It is important to note that the Lyapunov exponents are complex numbers. As is clear from Eq.~\eqref{growing-oscillations}, a generic complex $\zeta$ with positive real and non-vanishing imaginary parts indicates that the deviation from the AS solution shows an oscillatory behavior with a growing amplitude, where $\operatorname{Re}(\zeta)$ determines the growth rate while $\operatorname{Im}(\zeta)$ is the angular frequency of the oscillations. Let us now focus on the eigenvalue with the greatest real part, $\zeta_1$, which corresponds to the fastest-growing mode. This mode also has a nonvanishing imaginary part. In Figs.~\ref{fig-zeta-on-xi}(a) and (b), we show the real and imaginary parts of this root as functions of $\xi_0$. For each $\xi_0$, we actually present a band of values assumed by this root. The different values inside the band are obtained for different values of the axion--vector coupling in the range $\beta\in [10^{1.5},\,10^{3.5}]$. The blue bands shown in the two panels are obtained using the LGEF system while the green bands follow from Eq.~\eqref{eq-zeta-approx}. We find that the exact numerical results in blue are in excellent qualitative agreement and good quantitative agreement with the approximate analytical results in green. This observation serves as a validation and refinement of the results presented in Ref.~\cite{Peloso:2022ovc} and is one of the main results of the present work. The AS solution is unstable and the fastest-growing perturbation mode is characterized by the growth rate $\operatorname{Re}(\zeta_1)$ and oscillation frequency $\operatorname{Im}(\zeta_1)$ in Fig.~\ref{fig-zeta-on-xi}.

$\operatorname{Re}(\zeta_1)$ is a monotonically increasing function of $\xi_0$ and a decreasing function of $\beta$ at fixed $\xi_0$. The imaginary part exhibits the opposite behavior. For comparison, in Fig.~\ref{fig-zeta-on-xi}(b), we also show the estimate for the oscillation frequency found in Ref.~\cite{Domcke:2020zez},
\begin{equation}
\label{estimate-Domcke}
\omega=\operatorname{Im}(\zeta_1)\simeq \frac{\pi}{\ln(\xi_0^2/2)}.
\end{equation}
This expression follows from the fact that the response of the gauge-field Pontryagin density $\langle\bm{E}\cdot\bm{B} \rangle$ to changes in the inflaton velocity is retarded by approximately $\Delta N_\xi\simeq \ln(\xi_0^2/2)$. The main reason for this delay is that the growth rate of modes that cross the horizon and undergo the tachyonic instability at a certain moment of time is determined by the instantaneous value of $\xi$; however, these modes are still not dominating the spectrum of $\langle\bm{E}\cdot\bm{B} \rangle$ at this moment of time and will only do so $\Delta N_\xi$ $e$-folds later. We point out the good qualitative agreement between the estimate~\eqref{estimate-Domcke} and our numerical results. At the quantitative level, the two values agree up to a factor of roughly $1.5$.

\subsubsection{Decay of the AS solution}

Next, let us discuss what we shall refer to as the survival time of the AS solution. This quantity can be naturally defined as the moment of time when the relative deviation of the full numerical solution from the constant AS solution becomes of order unity. To be specific, in this work, we define the survival time as the first moment of time (or the number of $e$-folds from the beginning) when the production parameter $\xi$ deviates from the initial value in the AS solution, $\xi_0$, by more than half an order of magnitude,
\begin{equation}
\label{survival-time}
 N_{\mathrm{AS}} \equiv \min\limits_{i}N_i\,, \qquad \frac{|\xi(N_i)-\xi_0|}{\xi_0}=10^{-1/2} \,,
\end{equation}
where the $N_i$ denote the moments (in terms of the number of $e$-folds $N$) when the condition on the right-hand side is satisfied, and where we initialize the system at $N = 0$.

We emphasize that, in contrast to the Lyapunov exponents $\zeta_n$, which are intrinsic and characteristic properties of the physical system, the survival time $N_{\mathrm{AS}}$ depends on the way in which one imposes initial conditions.
For the sake of definiteness, we assume that all gauge-field bilinear quantities are in the AS solution corresponding to a certain value of the production parameter $\xi_0$, while the initial $\xi$ value is detuned to $\xi_0 + \delta\sm\xi_0$ by some small amount $\delta\sm\xi_0$. Since, for the constant background case considered in this section, the AS solution is an unstable equilibrium solution, the survival number of $e$-folds depends on the detuning parameter $\delta\sm\xi_0$ (it is infinite for $\delta\sm\xi_0 = 0$) as well as on the instability rate controlled by the Lyapunov coefficients studied above. 
As mentioned above, in the linear regime the quantity $\xi \left( N \right) - \xi_0$ is the linear superposition of a series of eigenmodes, each characterized by a Lyapunov exponent $\zeta_n$. Let us denote this series as 
\begin{equation}
\left\vert \xi \left( N \right) - \xi_0 \right\vert = \sum_n r_n \, \delta\sm\xi_0 \, {\rm Re } \left( {\rm e}^{\zeta_n N +i\varphi_n} \right)\;, 
\end{equation}
where the real coefficients $r_n$ and phases $\varphi_n$ depend on how the initial $\delta \sm\xi_0$ projects on each eigenmode, and where the initial number of $e$-folds has been set to $N=0$. Assuming a non-zero overlapping with the fastest growing mode (namely, the mode whose Lyapunov coefficient, denoted above as $\zeta_1$, has the greatest real part), and ignoring the initial phase $\varphi_1$, we then have 
\begin{equation}
\left\vert \xi \left( N \right) - \xi_0 \right\vert \simeq r_1 \, \delta\sm\xi_0 \, {\rm e}^{{\rm Re } \left( \zeta_1 \right) N} \;, 
\end{equation}
leading to the survival number of $e$-folds 
\begin{equation}
N_\mathrm{AS} \simeq - \frac{1}{{\rm Re } \left( \zeta_1 \right)} \left[ {\rm ln } \left\vert \frac{\delta\sm\xi_0}{\xi_0} \right\vert + \ln \left(10^{1/2} r_1 \right) \right] \simeq - \frac{1}{{\rm Re } \left( \zeta_1 \right)} \,  {\rm ln } \left\vert \frac{\delta\sm\xi_0}{\xi_0} \right\vert \, ,
\label{eq:NASestimate-v2}
\end{equation} 
where we ignore the order one term proportional to $\ln \left(10^{1/2} r_1 \right)$ in our discussion. We verify the dependence of $N_\mathrm{AS}$ on the initial conditions numerically in Appendix~\ref{app:survival-time} and show that it is in a good accordance with the estimate~\eqref{eq:NASestimate-v2}; see Fig.~\ref{fig-survival-time}. This provides a nontrivial check of the validity of our numerical scheme. Moreover, we study the dependence of the survival time on the parameters of the axion inflation model, which is also presented in Appendix~\ref{app:survival-time}. In this appendix, we also provide more details on how we choose the GEF initial conditions.

\begin{figure}
\centering
\includegraphics[width=0.48\linewidth]{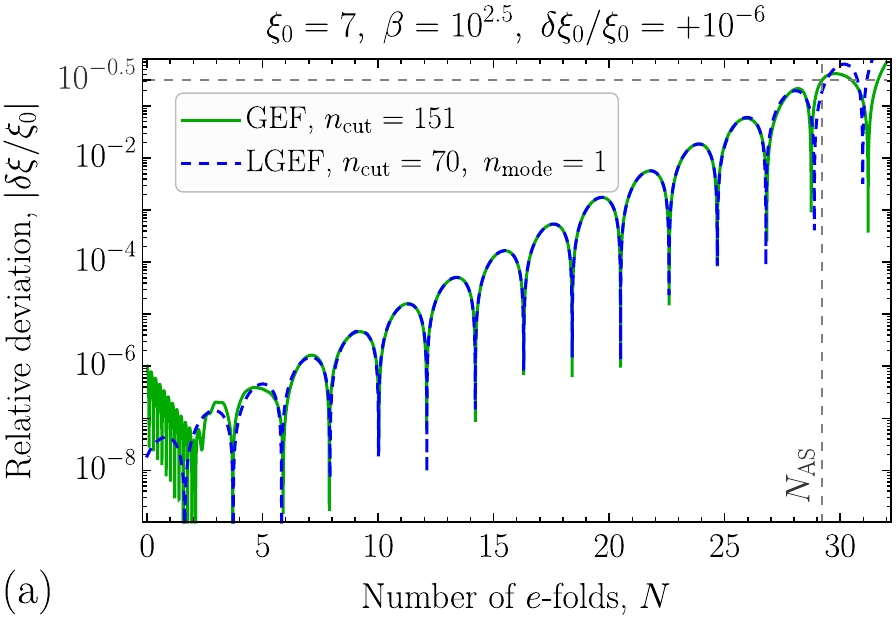}
\hspace{3mm}
\includegraphics[width=0.48\linewidth]{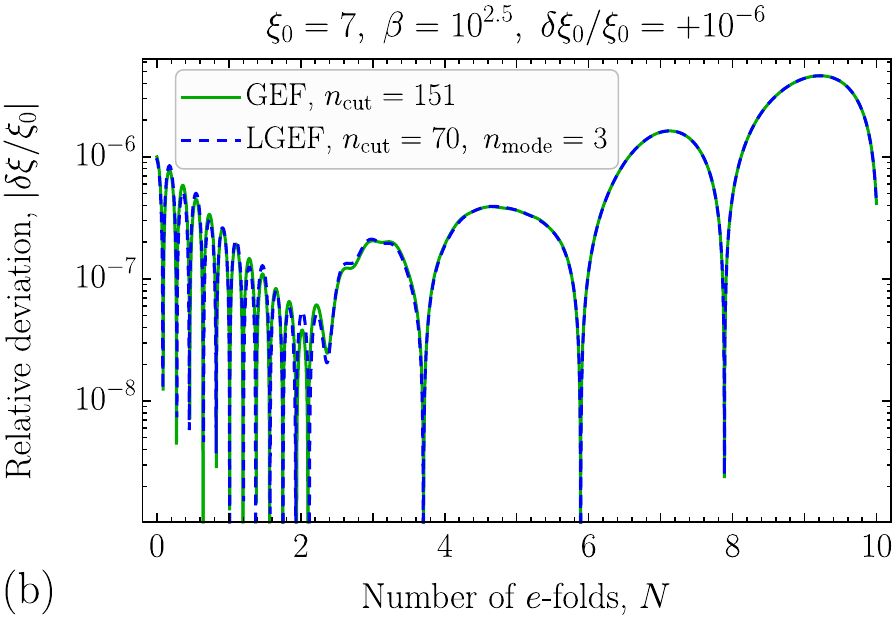}
\caption{Time evolution of the relative deviation $\delta\sm\xi/\xi_0$ for an initial $\xi$ value in the AS solution $\xi_0=7$, axion--vector coupling $\beta=10^{2.5}$, and initial fine-tuning (initial displacement) $\delta\sm\xi_0/\xi_0=10^{-6}$. (a)~Full time interval, (b)~zoom-in into the transition region from the initial decaying phase to the growing phase. The green solid lines show the solution obtained by the GEF truncated at order $n_\mathrm{cut}=151$; the blue dashed lines correspond to the result of the LGEF truncated at order $n_{\mathrm{cut}}=70$. In panel (a), only the fastest-growing mode $\zeta_1$ is included; in panel (b), three modes, $\zeta_{1,2,5}$, are included. The vertical gray dashed line in panel (a) indicates the survival time of the AS solution when the relative deviation of $\xi$ from the AS value $\xi_0$ reaches $10^{-1/2}$ (horizontal dashed line).}
\label{fig-evolution}
\end{figure}

In Fig.~\ref{fig-evolution}, we instead present one specific example of the departure from the AS solution. The evolution shown is characterized by $\beta=10^{2.5}$, $\xi_0=7$, and an initial fine-tuning of $\delta\sm\xi_0/\xi_0=10^{-6}$. The green solid line in the figure shows the evolution of $\delta\sm\xi/\xi_0$ as computed using the GEF system.
In panel (a) of Fig.~\ref{fig-evolution}, the evolution is tracked until the deviation becomes of order unity. The survival number of $e$-folds $N_\mathrm{AS}\approx 29$ is shown by the vertical gray dashed line. In panel (b), only the first 10 $e$-folds are shown. The blue dashed lines correspond to the LGEF solution (truncated at $n_\mathrm{cut}=70$) taking into account, respectively, only the fastest-growing mode in panel (a), and three eigenmodes $\zeta_1$, $\zeta_2$, and $\zeta_5$ in panel (b) (we recall that the eigenmodes are sorted by decreasing real part of $\zeta$). 

The first two eigenvalues, $\zeta_1$ and $\zeta_2$, have positive real parts, namely, they correspond to unstable departures from AS. The remaining eigenvalues have negative real parts. They correspond to stable departures of decreasing amplitudes. As visible from panel (a), the most unstable mode $\zeta_1$ is by itself able to account for the departure of $\xi$ from its initial value throughout the linearized stage. The inclusion of the other unstable modes, and, particularly, of the stable mode $\zeta_5$, well reproduces also the initial phase visible in the figure in which $\delta\sm\xi$ decreases.~\footnote{The eigenvalue $\zeta_5$ corresponds to the point in the upper half of Fig.~\ref{fig-complex-plane} that is well separated from the regular sequence of roots in the lower part of the plot. Its imaginary part is significantly greater than that of the first eigenvalues of the lower sequence. Correspondingly, $\delta\sm\xi$ oscillates much faster during the initial decreasing stage than in the following unstable phase. In passing, we also mention that a separated root, such as the $\zeta_5$-mode in the present benchmark scenario, is not always present in the spectrum. For instance, fixing $\beta = 10^{2.5}$ as the value considered in the figure, this separated root also exists for $\xi_0=5$, but not for $\xi_0=9$. The complicated form of the equations that we are solving, even in the simpler approximate form~\eqref{eq-zeta-approx}, does not allow us to determine a priori whether this separate root is present or not.} We stress that the decreasing stage visible in the figure is due to the fact that our choice of initial conditions is mostly mapped to this stable mode of eigenvalue $\zeta_5$, and it is not indicating that the AS solution is initially stable.

\begin{figure}
\centering
\includegraphics[width=0.48\linewidth]{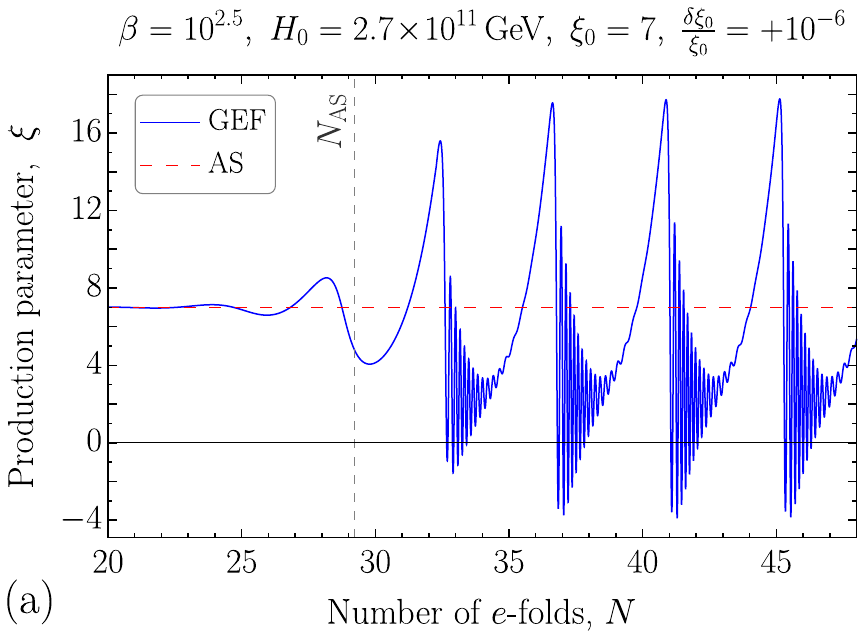}
\hspace{3mm}
\includegraphics[width=0.48\linewidth]{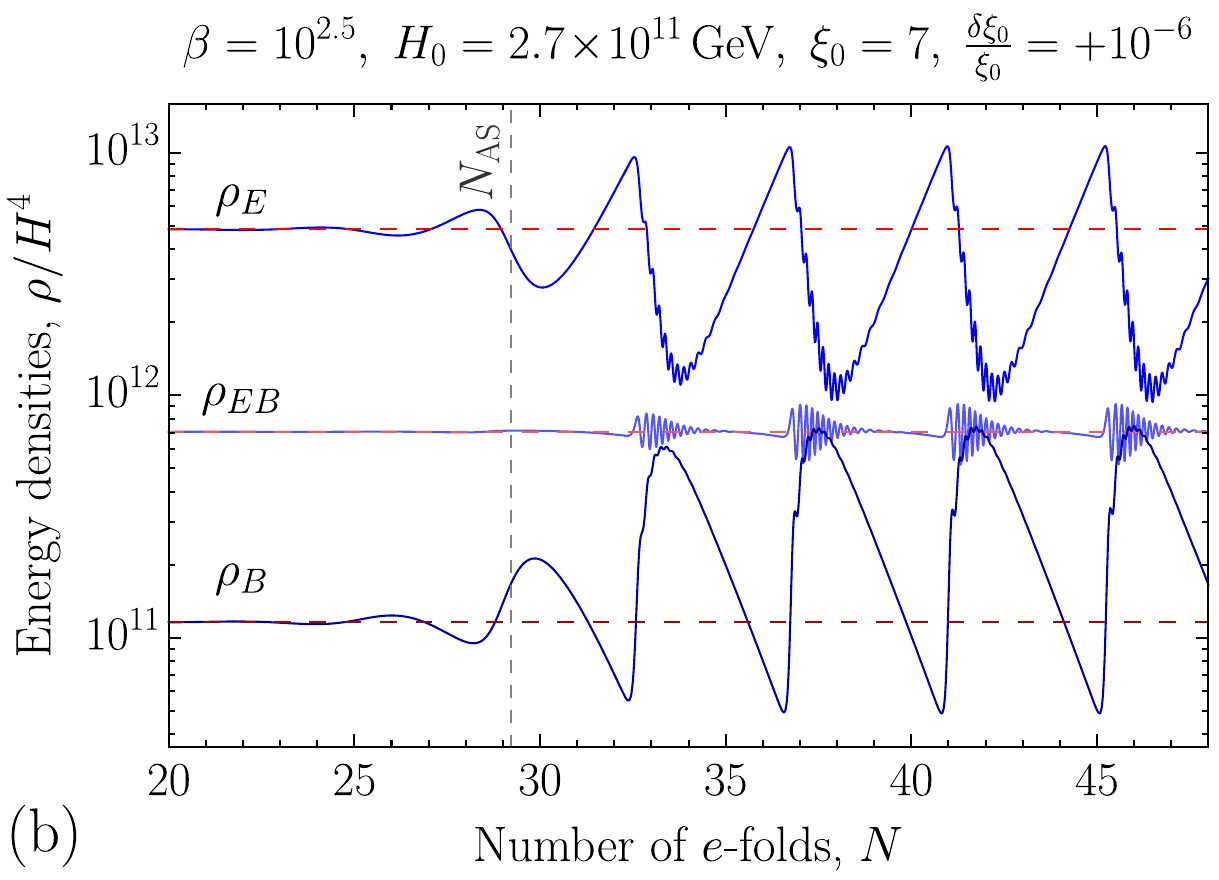}
\caption{Late-time evolution of (a)~the production parameter $\xi$ and (b)~energy densities $\rho_E=\langle\bm{E}^2\rangle/2$, $\rho_B=\langle\bm{B}^2\rangle/2$, and the Chern--Pontryagin density $\rho_{EB}=|\langle\bm{E}\cdot\bm{B}\rangle|/2$ for $\xi_0=7$, $\beta=10^{2.5}$, and $\delta\sm\xi_0/\xi_0=10^{-6}$. The vertical dashed lines show the moment of time $N=N_{\mathrm{AS}}$ when the linear perturbation theory breaks down (relative deviation of $\xi$ from the AS value exceeds $10^{-1/2}$).}
\label{fig-late-time}
\end{figure}

Finally, let us consider the late-time behavior of the solution, in the regime of large deviations from the AS solution, \textit{i.e.}, for $N>N_{\mathrm{AS}}$. This is the region where neither linear perturbation theory nor the LGEF are applicable. Therefore, we can only use the GEF in order to find the solution of the equations of motion. For the benchmark scenario with $\xi_0=7$ and $\beta=10^{2.5}$, the GEF solution is shown by the blue solid lines in Fig.~\ref{fig-late-time}. Panel (a) shows the time evolution of the $\xi$ parameter, while panel (b) illustrates the behavior of the produced gauge-field energy densities $\rho_E=\langle\bm{E}^2\rangle/2$, $\rho_B=\langle\bm{B}^2\rangle/2$, and the Chern--Pontryagin density $\rho_{EB}=|\langle\bm{E}\cdot\bm{B}\rangle|/2$. The red dashed lines show the corresponding AS solution. As we see from the plots, the time behavior of all quantities becomes almost perfectly periodic showing a sequence of highly oscillatory phases. In the simple case of constant background quantities, these oscillations will last indefinitely.

It is important to note that, for such a complicated and nonmonotonic behavior of the $\xi$ parameter as shown in Fig.~\ref{fig-late-time}, the cutoff momentum $k_\mathrm{h}$, for which Eq.~\eqref{k-h-final} gives
\begin{equation}
\label{k-h-final-N}
k_{\mathrm{h}}\left(N\right)=\max\limits_{N'\leq N}\left[2He^{N'}\left|\xi(N')\right|\right] \,,
\end{equation}
is not growing at all times; instead, there is a sequence of plateaus in $k_{\mathrm{h}}\left(N\right)$ where it remains constant. During these plateau stages, the truncation condition in Eq.~\eqref{truncation} may not be accurately satisfied because some of the underlying assumptions [see the paragraph below Eq.~\eqref{truncation}] are not valid at this time. In particular, the assumption that the spectral densities of $\mathcal{E}^{(n)}$, $\mathcal{G}^{(n)}$, and $\mathcal{B}^{(n)}$ are dominated by the mode $k_{\mathrm{h}}$ at large $n$ may be violated. These effects can lead to the breakdown of the GEF in a way that we discuss in Appendix~\ref{app:self-correction}. In this appendix, we also show one possible solution to this problem, namely, the GEF self-correction procedure.


\section{Time-dependent background quantities} 
\label{sec:realistic}

In this section we turn to the realistic case in which the Hubble parameter is not a constant, as assumed in the previous section, but it is consistently determined by the Friedmann equation \eqref{Friedmann}. For simplicity, we still consider the potential gradient $V'$ to be constant, meaning that the inflaton potential is a simple first-order polynomial,
\begin{equation}
\label{potential-constant-slope}
V(\phi)=V_0+ V' \phi.
\end{equation}
Such a potential is still not a realistic choice that could describe the whole stage of inflation over a large range in field space. A linear potential violates, \textit{e.g.}, the constraints on the tensor-to-scalar ratio imposed by CMB observations~\cite{Planck:2018jri,BICEP:2021xfz}, and does not allow for a graceful exit from the inflationary stage. Nevertheless, it serves as a good local approximation for a variety of potentials in restricted regions of field space. In any case, in the following, we shall use the simple potential in Eq.~\eqref{potential-constant-slope} primarily for illustrative purposes.

Firstly, let us give a recipe to determine the AS solution in any realistic inflationary model. As discussed in Sec.~\ref{sec:toolkit}, the AS solution ignores the retardation of the gauge-field response to the changes in the inflaton velocity. This allows to get a closed set of equations \eqref{KGF-AS-2}--\eqref{Friedmann-AS-2} determining the inflaton evolution. Since in the derivation of those equations we used the expressions for the gauge-field energy densities and the Chern--Pontryagin density for constant $\xi$ and $H$ parameters, the solution of Eqs.~\eqref{KGF-AS-2}--\eqref{Friedmann-AS-2} is not a solution of the full system of equations, which treats the Maxwell equations for the gauge field on the same footing as the Klein--Gordon and Friedmann equations. For this reason, we refer to the solution of Eqs.~\eqref{KGF-AS-2}--\eqref{Friedmann-AS-2} as the ``enforced'' AS (EAS) solution.

In order to find the numerical EAS solution of Eqs.~\eqref{KGF-AS-2}--\eqref{Friedmann-AS-2}, it is more convenient to rewrite the EOMs as a system of two first-order differential equations for the functions $\phi(t)$ and $\xi(t)$. In order to do so, we first solve Eq.~\eqref{Friedmann-AS-2} with respect to $H$,
\begin{equation}
H(\phi,\,\xi)=\left[\frac{2\,V(\phi)/M_{\mathrm{P}}^2}{3-2\,\xi^2/\beta^2+\sqrt{\left(3-2\,\xi^2/\beta^2\right)^2-2\left[e_0(\xi)+b_0(\xi)\right]V(\phi)/M_{\mathrm{P}}^4}}\right]^{1/2} \,.
\end{equation}
Then, the EOM for $\phi$ follows from the definition of $\xi$ in Eq.~\eqref{xi}, while the equation for $\xi$ can be derived from Eq.~\eqref{KGF-AS-2}. Finally, the desired system of equations has the form
\begin{align}
\label{eq-for-phi}
& \dot{\phi} = \frac{2M_{\mathrm{P}}}{\beta}H\left(\phi,\,\xi\right)\xi \,, \\
\label{eq-for-xi}
& \left(1+\xi\,\frac{\partial \ln H}{\partial\xi}\right)\dot{\xi}+\left(1+\frac{2\xi M_{\mathrm{P}}}{3\beta}\frac{\partial \ln H}{\partial\phi}\right)3H\xi=-\frac{\beta V'}{2HM_{\mathrm{P}}}-\frac{\beta^2 H^3}{2M_{\mathrm{P}}^2}\,g_0(\xi) \,.
\end{align}

\begin{figure}[p!]
\centering
\includegraphics[height=0.33\linewidth]{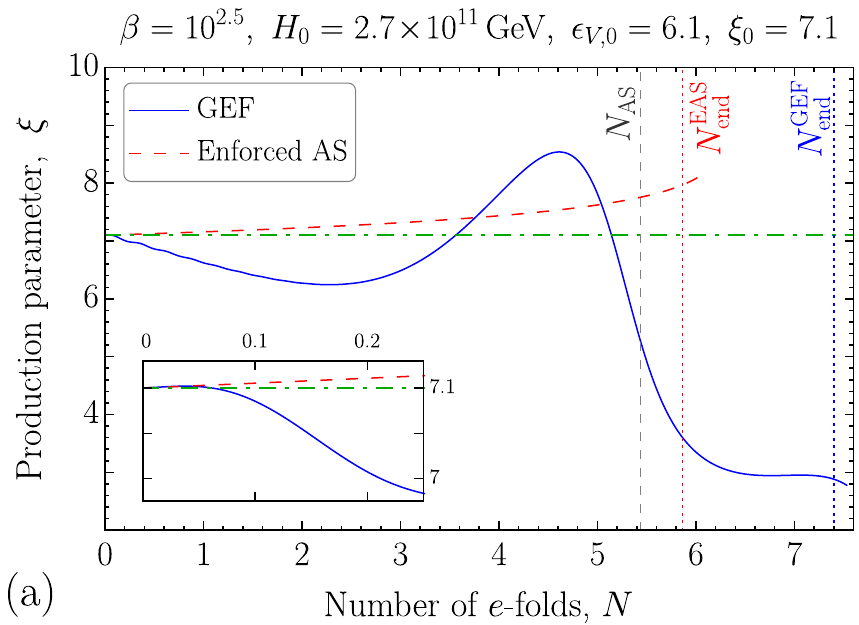}
\hspace{3mm}
\includegraphics[height=0.33\linewidth]{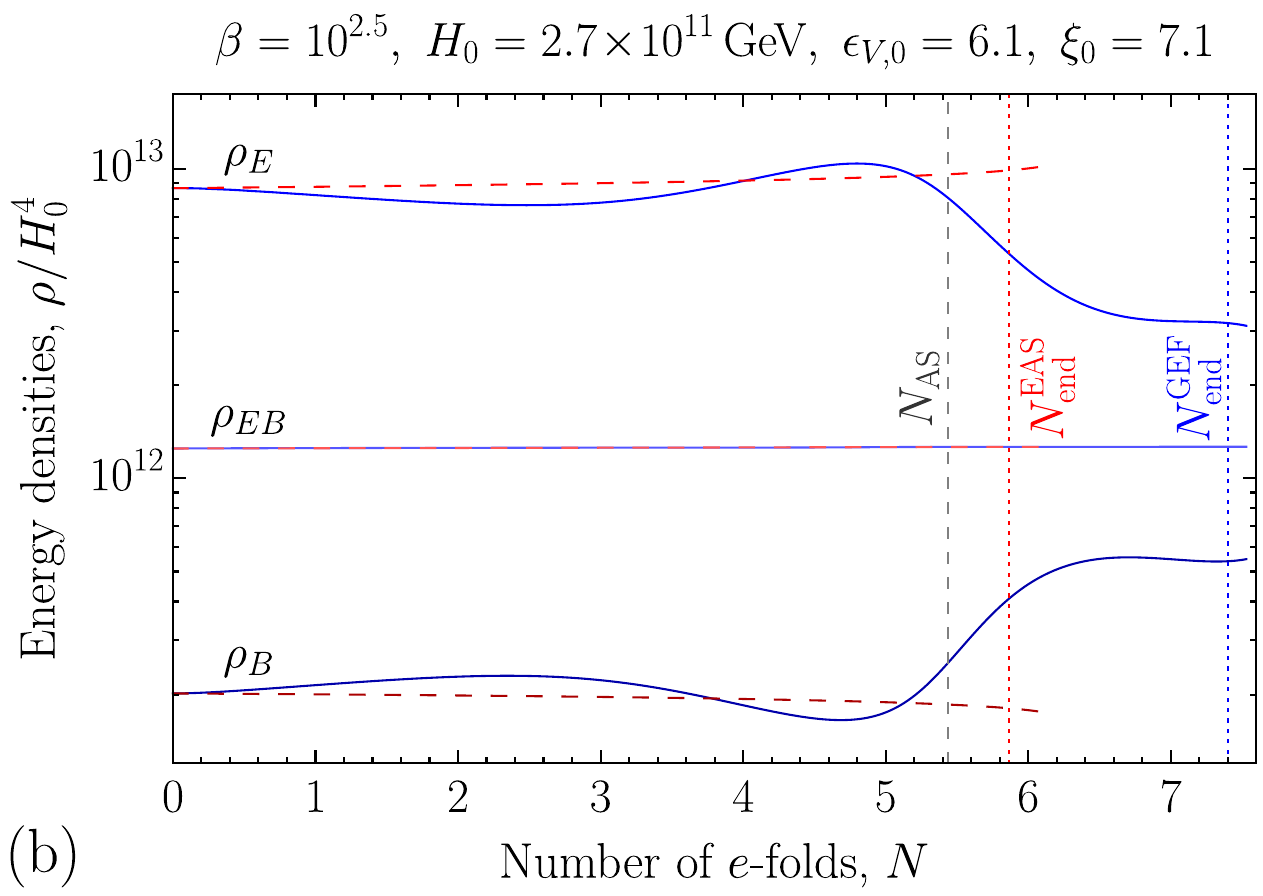}
\caption{Time evolution of (a)~the parameter $\xi$ and (b)~the energy densities for the axion--vector coupling $\beta=10^{2.5}$ in the realistic inflationary model featuring a steep inflaton potential with initial values of the Hubble parameter $H_0=2.7\times 10^{11}\,$GeV and slow-roll parameter $\epsilon_{V,0}=6.1$ (corresponding to $\xi_0=7.1$). The blue solid lines show the true solution of the system found using the GEF, while the red dashed lines correspond to the enforced AS solution. The vertical dashed lines show the moment of time $N=N_{\mathrm{AS}}$ when the linear perturbation theory breaks down (relative deviation of $\xi$ from its initial value exceeds $10^{-1/2}$). The vertical dotted lines show the end of inflation for the real system (blue) and for the enforced AS solution (red).}
\label{fig-SR-7-1}
\end{figure}

\begin{figure}[p!]
\centering
\includegraphics[height=0.33\linewidth]{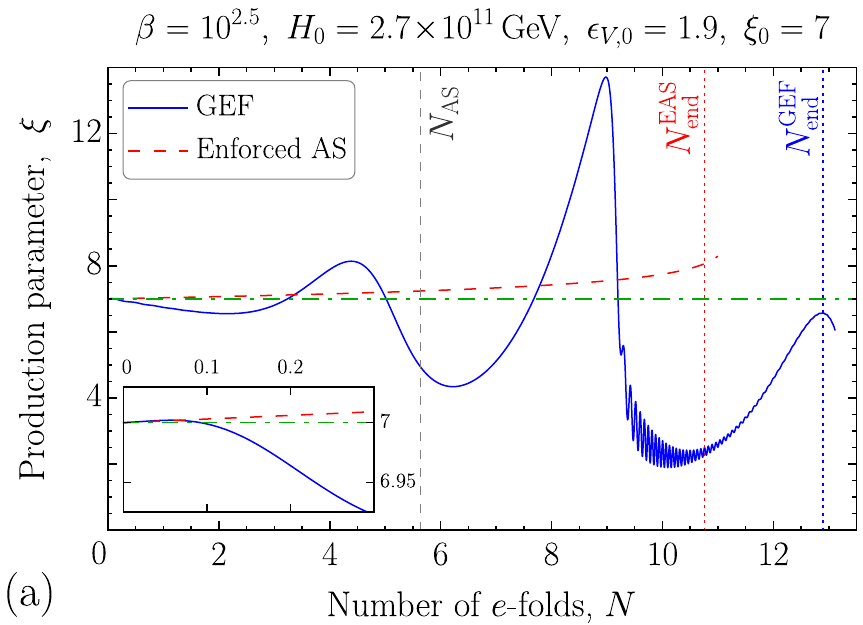}
\hspace{3mm}
\includegraphics[height=0.33\linewidth]{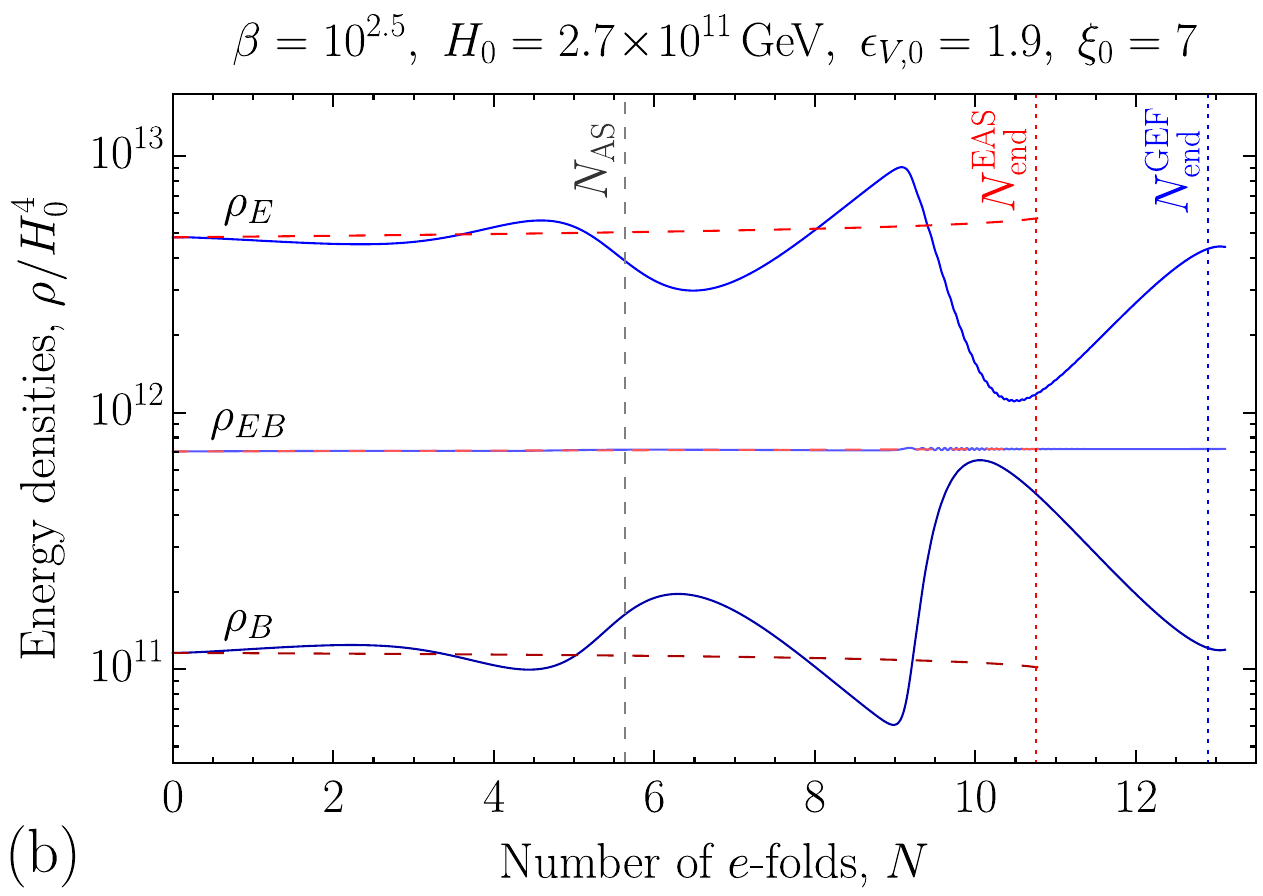}
\caption{Same as Fig.~\ref{fig-SR-7-1} but for initial slow-roll parameter $\epsilon_{V,0}=1.9$ (corresponding to $\xi_0=7$).}
\label{fig-SR-7}
\end{figure}

\begin{figure}[p!]
\centering
\includegraphics[height=0.33\linewidth]{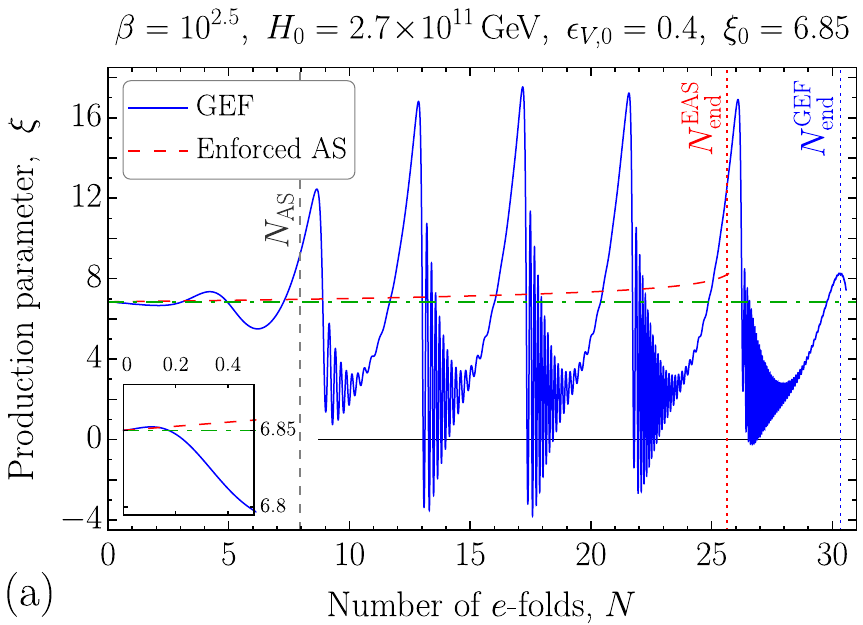}
\hspace{3mm}
\includegraphics[height=0.33\linewidth]{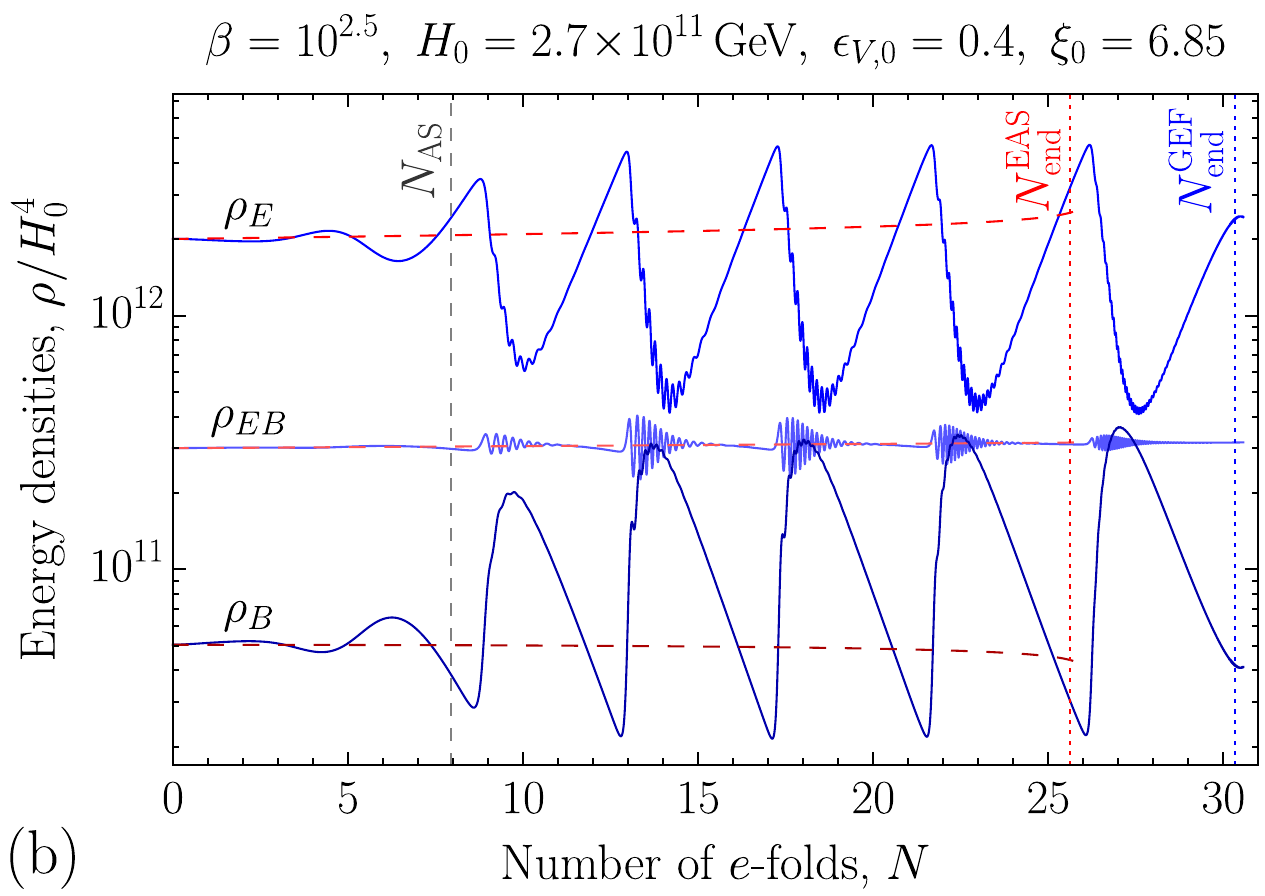}
\caption{Same as Fig.~\ref{fig-SR-7-1} but for initial slow-roll parameter $\epsilon_{V,0}=0.4$ (corresponding to $\xi_0=6.85$).}
\label{fig-SR-6-85}
\end{figure}

In order to study the stability of the AS solution in a fully time-dependent background, we have to initialize the system of Eqs.~\eqref{eq-for-phi}--\eqref{eq-for-xi} and, simultaneously, the full system of Klein--Gordon \eqref{KGF-2} and GEF equations \eqref{dot_E_n}--\eqref{dot_B_n} using \textit{exactly the same} initial conditions. We use the following algorithm to impose these initial conditions: 
\begin{itemize}
\item[(i)] Specify the axion--vector coupling $\beta$ and the desired value of $\xi_0$; propose a Hubble rate $H_0$ according to Eq.~\eqref{H-optimal} that puts the system in the strong-backreaction regime.
\item[(ii)] Use Eqs.~\eqref{KGF-AS-2}--\eqref{Friedmann-AS-2} to determine $V_0$ and $V'$ in the inflaton potential \eqref{potential-constant-slope}. In doing so, set $\ddot{\phi}_0 = 0$ and work with initial inflaton field value $\phi_0=0$, for definiteness.
\item[(iii)] Initialize the EAS system of Eqs.~\eqref{eq-for-phi}--\eqref{eq-for-xi} at a slightly smaller value of $\xi<\xi_0$ and $\phi<0$, to allow the system to cope with any potential inconsistency of the initial conditions. That is, let the system dynamically roll into a smooth and stable solution.
\item[(iv)] Determine the moment of time when the $\xi$ parameter in the EAS solution crosses the desired value $\xi_0$ and use it as the initial moment of time for the GEF. Note that this typically does not happen exactly at the origin of field space. We therefore no longer intend to initialize the system when $\ddot{\phi}_0 = 0$ and $\phi_0=0$. These conditions were used in step (ii) only to obtain some reasonable values for $V_0$ and $V'$. For given values of $V_0$ and $V'$, we can now forget about the fact that they were derived assuming $\ddot{\phi}_0 = 0$ and $\phi_0=0$. Instead, we now use the set of consistent values of $\phi$, $\ddot{\phi}$, and $H$ that we dynamically reach when the system has rolled into a smooth solution and $\xi$ corresponds to the desired value $\xi_0$.
\footnote{In fact, we find that this approach results in a numerically negligible deviation between $H$ and $H_0$, less than $0.01\,\%$. In Figs.~\ref{fig-SR-7-1}--\ref{fig-SR-6-85} discussed below, we therefore label all plots with the $H_0$ value chosen in step (i), even though the self-consistent value of $H$ is only determined in step (iv).}
The advantage of this procedure is that it allows us to initialize the GEF with a self-consistent set of input values, including $\ddot{\phi}$ and hence the time derivative of $\xi$.
\item[(v)] Use $\phi$ and $\dot{\phi}$ from the previous step to initialize the Klein--Gordon equation \eqref{KGF-2} for the GEF; compute the initial conditions for the bilinear functions in the GEF by inserting $\xi_0$ and $H$ into Eqs.~\eqref{IC-En}--\eqref{IC-Bn} in Appendix~\ref{app:bilinears}.
\end{itemize}
Below, we present the numerical results we obtain using the GEF and compare them to the corresponding results based on the EAS solution.

Figures~\ref{fig-SR-7-1}--\ref{fig-SR-6-85} present the results of the evolution for three different realizations of the model. More precisely, in all three figures the same value of the axion--vector coupling $\beta=10^{2.5}$ and of the Hubble parameter $H_0=2.7\!\times\!10^{11}\,$GeV are assumed. The three figures differ by a decreasing steepness of the potential $V'$ and, consequently, in the correspondingly required value of $\xi_0$ (the specific values of $\xi_0$ and of the slow-roll parameter $\epsilon_{V,0}$ assumed in each figure are shown in their titles). Panels (a) of each figure show the evolution of the $\xi$ parameter as a function of number of $e$-folds $N$, while panels (b) of each figure show the evolution of the energy densities $\rho_E$, $\rho_B$, and the Chern--Pontryagin density $\rho_{EB}$ of the produced gauge fields. In all figures and panels, the blue solid lines show the exact solution of the full system of equations obtained from the GEF, while the red dashed lines show the corresponding quantities in the EAS solution. Vertical dotted lines of the same color show the moment of time when inflation ends ($\ddot{a} = 0$) for the chosen initial conditions, and the gray dashed vertical lines denote as usual the survival time of the AS solution, \textit{i.e.}, the moment when the relative deviation of the blue curve from the red curve for the evolution of $\xi$ reaches the threshold value of $10^{-1/2}$ for the first time. In the following, we discuss a few features that can be read off from these figures.


Firstly, we note that the survival time of the EAS solution cannot be increased to infinity by fine-tuning the initial conditions. This is due to the fact that, once $H$ and $\dot{\phi}$ are allowed to consistently vary, the EAS solution is no longer an exact solution of the full system of equations. This poses an upper limit on the survival time, which we typically found to be around 5 to 7 $e$-folds. 
Secondly, the evolution of the exact numerical solution of the system is qualitatively the same as in the simple case of constant $H$  considered in Sec.~\ref{sec:const_back}. Indeed, the curve for the $\xi$ parameter initially shows oscillations around the AS solution with a growing amplitude until the deviations becomes eventually of order unity and the growth stops due to nonlinearities in the system. After that, a new phase typically sets in, characterized by quasiperiodic stages of fast oscillations. However, now, due to the continuously changing Hubble parameter, the amplitude and oscillation frequency slowly change in time.
Thirdly, the comparison between the different figures shows that the duration of inflation  strongly depends on the steepness of the potential. Note that, for the case of a very steep potential (with initial value of the slow-roll parameter $\epsilon_{V,0}\approx 6.1$) shown in Fig.~\ref{fig-SR-7-1}, the duration of inflation is just 7 $e$-folds. The second phase of fast oscillations of $\xi$ does not even start in this case. For a flatter potential with $\epsilon_{V,0}\approx 1.9$, shown in Fig.~\ref{fig-SR-7}, inflation lasts for approximately 13 $e$-folds, allowing for one stage of fast oscillations. Further flattening the potential with $\epsilon_{V,0}\approx 0.4$, as shown in Fig.~\ref{fig-SR-6-85}, the duration of inflation is greater than 30 $e$-folds, and here five periods of fast oscillations appear.%
\footnote{In the case $\epsilon_{V,0}\approx 0.4$, the slow-roll inflation occurs even in the absence of gauge fields; however, it lasts less than one $e$-fold and is followed by hitting the value $V(\phi)=0$. The gauge fields not only extend the duration of inflation but also lead to the radiation-dominated universe shortly before the potential hits the zero value.}
Note that the fast oscillatory stages in the $\xi$ evolution lead again to the plateaus in the evolution of $k_{\mathrm{h}}$, which complicates the integration of the GEF system. In particular, in order to obtain the numerical results with a controllable accuracy until the end of inflation, we need to apply several self-correction procedures, which we discuss in Appendix~\ref{app:self-correction}.

\begin{figure}
\centering
\includegraphics[height=0.4\linewidth]{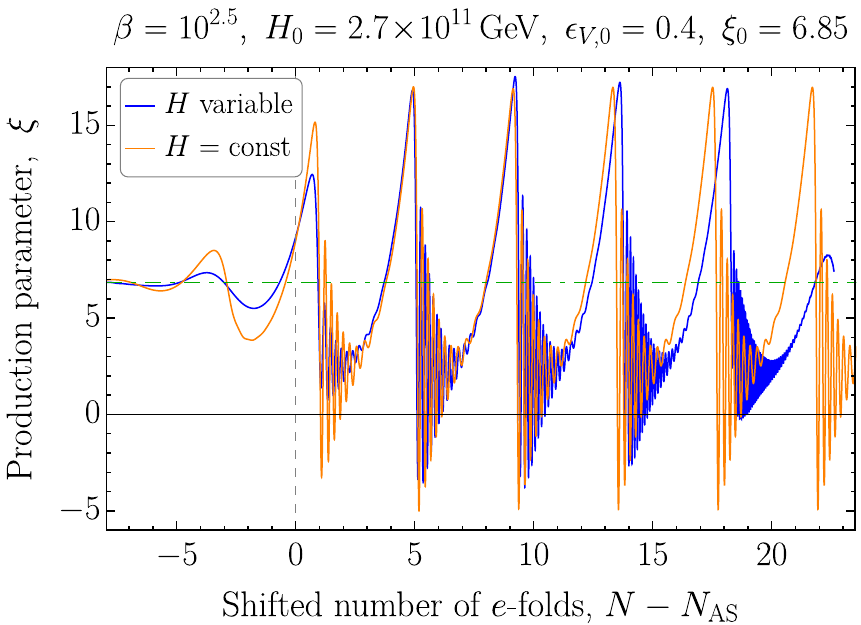}
\caption{Time evolution of the parameter $\xi$ in the realistic inflationary model (blue line) compared to the late-time behavior of $\xi$ in the toy model with constant $H$ (orange line) studied in Sec.~\ref{sec:const_back}. The curves are shifted in such a way that the moments of time when the condition $(\xi-\bar{\xi})/\bar{\xi}=+10^{1/2}$ is satisfied for the first time coincide (vertical dashed line), where $\bar{\xi}$ denotes either the value $\xi_0=6.85$ in the case of constant background quantities or the time-dependent $\xi$ value according to the EAS solution in the case of time-dependent background. The green dashed-dotted horizontal line shows the initial value $\xi_0=6.85$ for both curves.}
\label{fig-SR-6-85-compare}
\end{figure}

In Fig.~\ref{fig-SR-6-85-compare}, we further elaborate on the comparison of the evolution of the $\xi$ parameter obtained in a realistic inflationary model (the same blue solid line as in Fig.~\ref{fig-SR-6-85}) with that obtained under the assumption of constant $H$ (equal to the initial $H$ value in the realistic model; the evolution of $\xi$ in this case is shown by an orange solid line). Both evaluations are shown as a function of the number of $e$-folds $N$, with the value $N = N_{\rm AS}$ set when the $\xi$ parameter has increased by half an order of magnitude with respect to the corresponding AS value (which equals $\xi_0=6.85$ in the case of constant background quantities and which is taken from the EAS solution in the case of time-dependent background).
The figure clearly demonstrates that the two solutions are very close to each other, not only qualitatively but even showing a good quantitative agreement. In particular, the amplitude and frequency of the large-amplitude oscillations in the nonlinear-perturbation regime are in a good agreement. We only observe a slow drift of the parameters due to slow-roll corrections for the realistic inflationary model. This result means that the case of constant background quantities considered in great detail in Sec.~\ref{sec:const_back}, despite its simplicity, still allows to obtain a good intuition for the time evolution in the realistic case. We attribute this agreement to the fact that, also in the realistic case, $H$ changes only very slowly during inflation.


\section{Conclusions}
\label{sec:conclusion}

An axion-like inflaton coupled to a gauge field provides probably one of the simplest and best motivated models where matter production occurs during inflation. Several studies in the last few years have shown that, in the strong-backreaction regime, this model displays a highly nontrivial behavior, significantly different from the steady state AS solution. Such a behavior has been mostly studied via numerical techniques (with the exception of Ref.~\cite{Peloso:2022ovc}), and the origin and the fate of this  departures are the subject of ongoing research.

In this work, we used the gradient expansion formalism developed in Refs.~\cite{Sobol:2019xls,Gorbar:2021rlt,Durrer:2023rhc} to improve our understanding of the system. In the first part of our work, we studied a setup similar to that of Ref.~\cite{Peloso:2022ovc}, where the velocity of the inflaton is initially close to its AS value, under the assumption of a constant Hubble parameter. The analysis we presented in Sec.~\ref{sec:const_back} leads to results that agree with those of Ref.~\cite{Peloso:2022ovc}: among the complex Lyapunov exponents for perturbations around the AS solution, there is always at least one exponent with a positive real part, which results in oscillation of increasing amplitude of the inflaton velocity about the value predicted by the steady state AS solution. While the results of Sec.~\ref{sec:const_back} were obtained for a system identical to that of Ref.~\cite{Peloso:2022ovc}, the techniques used, and in particular the approximations, are significantly different. For instance, Ref.~\cite{Peloso:2022ovc} uses an approximate Green function, which is not required in the solution of Sec.~\ref{sec:const_back}, whereas the truncation of the hierarchy of equations in Sec.~\ref{sec:const_back} has no counterpart in the analysis of Ref.~\cite{Peloso:2022ovc}. For this reason, those results corroborate each other. In the second part of our work, we studied the onset of the instability in the case in which the Hubble parameter is evolving over time. Previous numerical studies in the literature started in the weak-backreaction regime and saw the instability building up as the backreaction becomes strong. One might therefore wonder whether the AS solution might still be stable (even if, possibly, with a very small basin of attraction) if one started directly from the strong-backreaction regime. The analysis of Sec.~\ref{sec:const_back}, which assumes $\dot{H}=0$, does not fully settle this question, since in that case $\delta\sm\xi=0$ is a valid (albeit unstable) solution that might in principle be stabilized when $\dot{H}\neq 0$. Although it is hard to imagine that a slow-roll variation of $H$ could change this behavior, this remained a logical possibility from the existing literature. Our analysis of Sec.~\ref{sec:realistic} shows that this is not the case. Even if we place our solution on the AS values in the strong-backreaction regime ``by hand'', the time dependence of the Hubble parameter will destabilize the system, leading in only a few $e$-folds to the oscillating behavior observed in previous studies of the transition from weak to strong backreaction.

While we believe that our work clarifies, and possibly settles, the questions around the onset of the instability in the axion--vector system, there are still open questions concerning the subsequent evolution and the possible end of the unstable regime. Thus far, these questions have been tackled only with the use of numerical studies. The majority of those studies assumes a spatially uniform inflaton field, which results in a quasi-periodic pattern of oscillations in the inflaton velocity, where deviations from perfectly periodic oscillations are only due to the slow-roll evolution of the inflaton zero mode probing different parts of the potential at different times. So far, only two works have considered the effects of spatial fluctuations of the inflaton field during the inflationary stage (and none has considered the effects of metric perturbations around an FLRW background). These lattice studies, being computationally expensive, covered only a relatively brief time interval. The study of Ref.~\cite{Caravano:2022epk} was able to capture the first oscillation of the inflaton velocity, whose shape agrees with that found in the works with a homogeneous inflaton. More recently, Ref.~\cite{Figueroa:2023oxc} studied the system for a more extended time range, showing that, around the time of the first oscillation, spatial inhomogeneities in the inflaton field build up very rapidly, and that the subsequent oscillations in the inflaton velocity have a suppressed amplitude. This is a relevant qualitative change with respect to the previous results, which warrants further study. How does this behavior depend on parameters? (Indeed, the pattern of oscillations in the examples shown in~\cite{Figueroa:2023oxc} changes significantly for very small variations of the axial coupling.) To what extent are the results in Ref.~\cite{Figueroa:2023oxc} affected by the fact that the simulations are probing only the last $\lesssim 10$ $e$-folds of inflation? We hope that a (semi-)analytical study extending the formalism presented here might shed more light on these questions; we plan to return to this problem in the future.


\vskip.25cm
\section*{Acknowledgements} 
M.\,P.\ and K.\,S.\ would like to thank the organizers of the Bethe Forum on Axions at the Bethe Center for Theoretical Physics in October 2022, where this project was initiated.
The authors are grateful to Valerie Domcke, Yohei Ema, Kyohei Mukaida, and Stefan Sandner for fruitful discussions. 
The work of R.\,v.\,E.\ and K.\,S.\ is supported by the Deutsche Forschungsgemeinschaft (DFG) through the Research Training Group, GRK 2149: Strong and Weak Interactions\,---\,from Hadrons to Dark Matter. M.\,P.\ is supported by Istituto Nazionale di Fisica Nucleare (INFN) through the Theoretical Astroparticle Physics (TAsP) project. O.\,S.\ is supported by a Philipp Schwartz fellowship of the University of M\"{u}nster and by the National Research Foundation of Ukraine Project No.2020.02/0062. The work of L.\,S.\ is partially supported by the US-NSF grant PHY-2112800.


\appendix
\section{Bilinear functions for constant \boldmath{$\xi$} and \boldmath{$H$}}
\label{app:bilinears}

In the case of constant $\xi$ and $H$, \textit{i.e.}, for a constant inflaton velocity in a stationary de Sitter spacetime, the gauge-field mode functions are given by the expression in Eq.~\eqref{A-Whittaker}, which enters the spectral representation of the bilinear electromagnetic functions,
\begin{align}
\mathcal{E}^{(n)} & = \int\limits_{0}^{k_{\mathrm{h}}}\frac{dk}{k}\frac{k^{n+3}}{2\pi^2 a^{n+4}}\sum\limits_{\lambda=\pm}\lambda^{n}|A'_\lambda(\tau,k)|^{2}, \label{En-sp}
\\
\mathcal{G}^{(n)} & = \int\limits_{0}^{k_{\mathrm{h}}}\frac{dk}{k}\frac{k^{n+4}}{2\pi^2 a^{n+4}}\sum\limits_{\lambda=\pm}\lambda^{n+1}\operatorname{Re}[A^{\ast}_\lambda(\tau,k)A'_\lambda(\tau,k)],\label{Gn-sp}
\\    
\mathcal{B}^{(n)} & = \int\limits_{0}^{k_{\mathrm{h}}}\frac{dk}{k}\frac{k^{n+5}}{2\pi^2 a^{n+4}}\sum\limits_{\lambda=\pm}\lambda^{n}|A_\lambda(\tau,k)|^{2},\label{Bn-sp}
\end{align}
where $k_{\mathrm{h}}=k_{\mathrm{h}}(\tau)=2\xi aH=-2\xi/\tau$. Substituting the explicit form of the mode functions into these integrals, one obtains the following expressions:
\begin{equation}
\mathcal{E}^{(n)} = H^{n+4} e_n(\xi) \,, \qquad \mathcal{G}^{(n)} = H^{n+4} g_n(\xi) \,, \qquad \mathcal{B}^{(n)} = H^{n+4} b_n(\xi) \,,
\end{equation}
where
\begin{align}
e_n\left(\xi\right) & = \int\limits_0^{2\xi} \frac{dx}{4\pi^2} \sum\limits_{\lambda=\pm}\lambda^{n} x^{n+1}e^{\lambda\pi\xi}\left|\left(x-\lambda\xi\right)W_{-i\lambda\xi,1/2}\left(-2ix\right)-iW_{1-i\lambda\xi,1/2}\left(-2ix\right)\right|^2 \,, \label{IC-En}\\
g_n\left(\xi\right) & = \int\limits_0^{2\xi}\frac{dx}{4\pi^2} \sum\limits_{\lambda=\pm}\lambda^{n+1} x^{n+2}e^{\lambda\pi\xi}\operatorname{Re}\left[W_{i\lambda\xi,1/2}\left(2ix\right)W_{1-i\lambda\xi,1/2}\left(-2ix\right)\right] \,, \label{IC-Gn}\\
b_n\left(\xi\right) & = \int\limits_0^{2\xi}\frac{dx}{4\pi^2} \sum\limits_{\lambda=\pm}\lambda^{n} x^{n+3}e^{\lambda\pi\xi}\left|W_{-i\lambda\xi,1/2}\left(-2ix\right)\right|^2 \,. \label{IC-Bn}
\end{align}
These expressions are valid for any value of $\xi>0$. However, in the case of strong gauge-field production when $\xi\gtrsim 5$, it is possible to derive simple approximate expressions for the bilinear functions taking into account that (i)~only one circular polarization, $\lambda=+$, is exponentially enhanced, while the second one can be safely neglected, and (ii)~there is a simple representation for the mode function in terms of elementary functions that is valid for $-k\tau \ll 2\xi$ [see Eq.~\eqref{A-plus-approx}]. Here, the second condition can be justified by the fact that the spectral densities for the lowest-order bilinear functions are maximal in the region $|k\tau|\sim 1/\xi$~\cite{Domcke:2020zez}, which is indeed much less than $2\xi$ for sufficiently large $\xi$. However, both conditions listed above are not valid for the bilinear quantities of high order $n$. For them, the spectral densities are peaked at the UV boundary of the spectrum, \textit{i.e.}, at $k\sim k_{\mathrm{h}}$, because of the additional $k^n$ factor in the integrals \eqref{En-sp}--\eqref{Bn-sp}. In this region, the mode functions of both polarizations are of the same order of magnitude and cannot be approximated by Eq.~\eqref{A-plus-approx}. Below, we state nonetheless the approximate expressions for the bilinear functions based on the mode functions in Eq.~\eqref{A-plus-approx} for arbitrary $n$. However, one must keep in mind that the applicability of these expressions is somewhat limited.

Neglecting the exponentially suppressed term in Eq.~\eqref{A-plus-approx}, we have the following expression for the mode function:
\begin{equation}
A_{+}\left(\tau,k\right) \simeq \frac{1}{\sqrt{2k}}\left(\frac{x}{2\xi}\right)^{1/4} e^{\pi\xi-2\sqrt{2\xi x}}, \qquad x=-k\tau \,.
\end{equation}
In its time derivative, one should differentiate only the exponential function since it is changing the fastest
\begin{equation}
A'_{+}\left(\tau,k\right) \simeq \sqrt{\frac{k}{2}}\left(\frac{x}{2\xi}\right)^{-1/4}e^{\pi\xi-2\sqrt{2\xi x}} \,.
\end{equation}
Then, we obtain the following results for the dimensionless bilinear functions:
\begin{equation}
e_n^{\mathrm{approx}}=\frac{(2n+6)!}{2^{5n+18}\pi^2} \frac{e^{2\pi\xi}}{\xi^{n+3}} \,,\quad g_n^{\mathrm{approx}}=\frac{(2n+7)!}{2^{5n+21}\pi^2} \frac{e^{2\pi\xi}}{\xi^{n+4}} \,,\quad b_n^{\mathrm{approx}}=\frac{(2n+8)!}{2^{5n+24}\pi^2} \frac{e^{2\pi\xi}}{\xi^{n+5}} \,.
\end{equation}
In order to impose the initial conditions for the GEF, we work with Eqs.~\eqref{IC-En}--\eqref{IC-Bn}. We only use $e_0^{\mathrm{approx}}$, $g_0^{\mathrm{approx}}$, and $b_0^{\mathrm{approx}}$ in our estimate of the parameter range in Sec.~\ref{subsec:parameters}.


\section{Survival time of the AS solution}
\label{app:survival-time}

In this Appendix, we discuss in more detail the way in which we impose the GEF initial conditions in Sec.~\ref{sec:const_back}, and we study the dependence of the survival time of the AS solution, $N_{\rm AS}$, on the initial conditions and the parameters of the axion inflation model in the toy model that assumes a constant Hubble rate $H$ throughout the whole evolution.

The exact procedure by which to perturb the initial conditions requires some thought, especially, in view of the fact that we intend to use the GEF to determine the full numerical solution of the system. The GEF system that we wish to integrate has $(3n_{\mathrm{cut}}+4)$ unknown functions, where $n_{\mathrm{cut}}$ is the truncation order of the system, and therefore there are \textit{plenty} of ways to perturb the initial conditions. To be concrete, we will opt for one of the simplest possibilities in our analysis, consisting of the following four-step algorithm:
\begin{itemize}
\item[(i)] For a given value of the axion--vector coupling constant $\beta$ in the Lagrangian, choose an initial value of the  parameter $\xi$ in the AS solution, $\xi_0$.
\item[(ii)] Use Eq.~\eqref{H-optimal} to determine the optimal Hubble rate (allowing us to realize $\delta_{\mathrm{KG}} \gg 1$ and $\delta_{\mathrm{F}} \ll 1$); find the corresponding gradient of the inflaton potential from Eq.~\eqref{AS-const}.
\item[(iii)] Compute the initial conditions for the bilinear quantities $\mathcal{E}^{(n)}$, $\mathcal{G}^{(n)}$, $\mathcal{B}^{(n)}$ according to Eqs.~\eqref{IC-En}--\eqref{IC-Bn} in Appendix~\ref{app:bilinears} for the values of $\xi_0$ and $H$ fixed in (i) and (ii).
\item[(iv)] Perturb the initial $\xi$ value by a small deviation from $\xi_0$, \textit{i.e.}, $\xi_0 \to \xi_0+\delta\sm\xi_0$. The relative deviation $\delta\sm\xi_0/\xi_0$ is a free parameter that controls the AS survival time.
\end{itemize}

\begin{figure}
\centering
\includegraphics[width=0.67\linewidth]{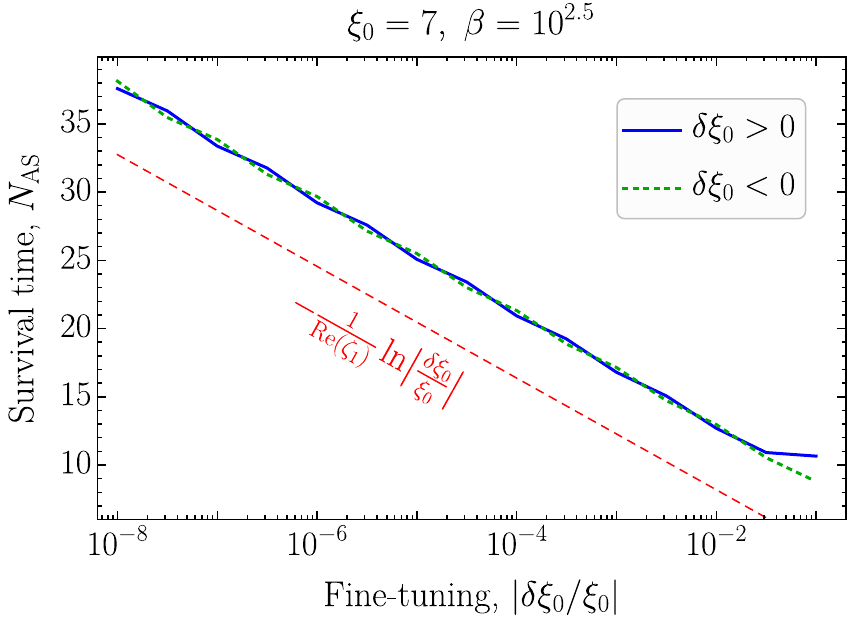}
\caption{Survival time of the AS solution as a function of the initial fine-tuning $\delta\sm\xi_0/\xi_0$ for $\xi_0=7$ and axion--vector coupling $\beta=10^{2.5}$. The blue solid and green dotted lines correspond to positive and negative initial deviations, respectively. The red dashed line shows the analytical estimate of the survival time in Eq.~\eqref{eq:NASestimate-v2}, which is based on the growth rate of the fastest-growing mode.}
\label{fig-survival-time}
\end{figure}

The survival time of the AS solution is determined according to Eq.~\eqref{survival-time} as the first moment of time when the relative deviation of the exact solution of the system deviates from the AS solution by half an order of magnitude. For example, for the axion--vector coupling $\beta=10^{2.5}$ and initial relative deviation $\delta\sm\xi_0/\xi_0=10^{-6}$ from the AS solution $\xi_0=7$, the survival time $N_\mathrm{AS}\approx 29$ $e$-folds, according to Fig.~\ref{fig-evolution} in the main text. If we choose other values for the fine-tuning of the initial condition, $\delta\sm\xi_0/\xi_0$, the survival time of the AS solution will change. This dependence is shown in Fig.~\ref{fig-survival-time}. The blue solid line (green dotted line) corresponds to the initial deviation in the direction of greater (smaller) $\xi$ values. 
Not surprisingly, the survival time has a clear logarithmic dependence on $\left\vert \delta\sm\xi_0 \right\vert$ as long as this quantity is in the linear regime, in agreement with Eq.~\eqref{eq:NASestimate-v2}. This general decreasing trend visible in Fig.~\ref{fig-survival-time} is easy to understand: the closer we are to the AS solution initially, the more time it will take the deviation to grow until they become of order unity. This dependence can be simply estimated as in Eq.~\eqref{eq:NASestimate-v2}, which is shown by the red dashed line in Fig.~\ref{fig-survival-time} and nicely reproduces the slope of the exact solution. The constant shift of this line can be explained by the fact that Eq.~\eqref{eq:NASestimate-v2} underestimates the survival time since it does not account for the initial decreasing stage.~\footnote{This initial stage exists because our initial deviation from the AS solution has a nonzero overlap with also some decreasing mode, see Figure~\ref{fig-evolution}.}  Deviations from this dependence (small wiggles on the blue and green lines) occur because of the phase of the cosine function when the solution crosses the threshold value $10^{-1/2}$. The agreement with the analytical results confirms the robustness of our numerical techniques also for the small departures shown in the figure.

\begin{figure}[ht]
\centering
\includegraphics[width=0.48\linewidth]{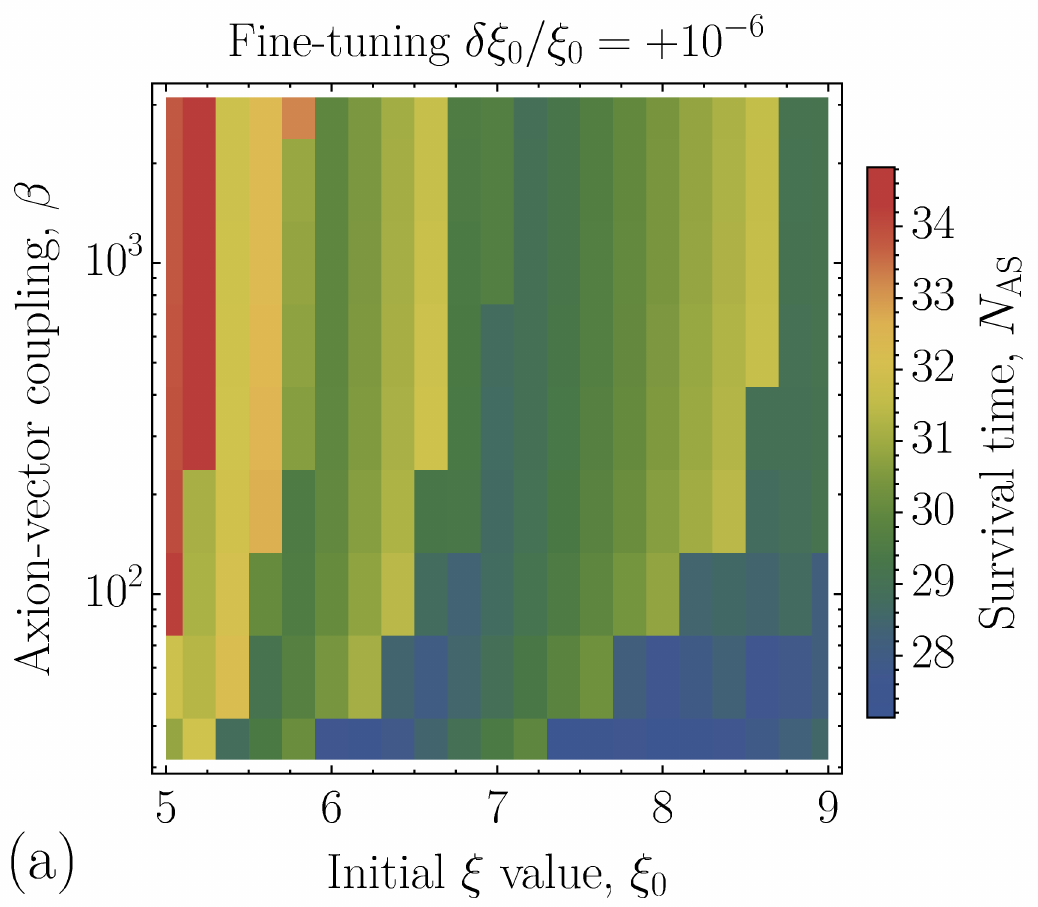}
\hspace{3mm}
\includegraphics[width=0.48\linewidth]{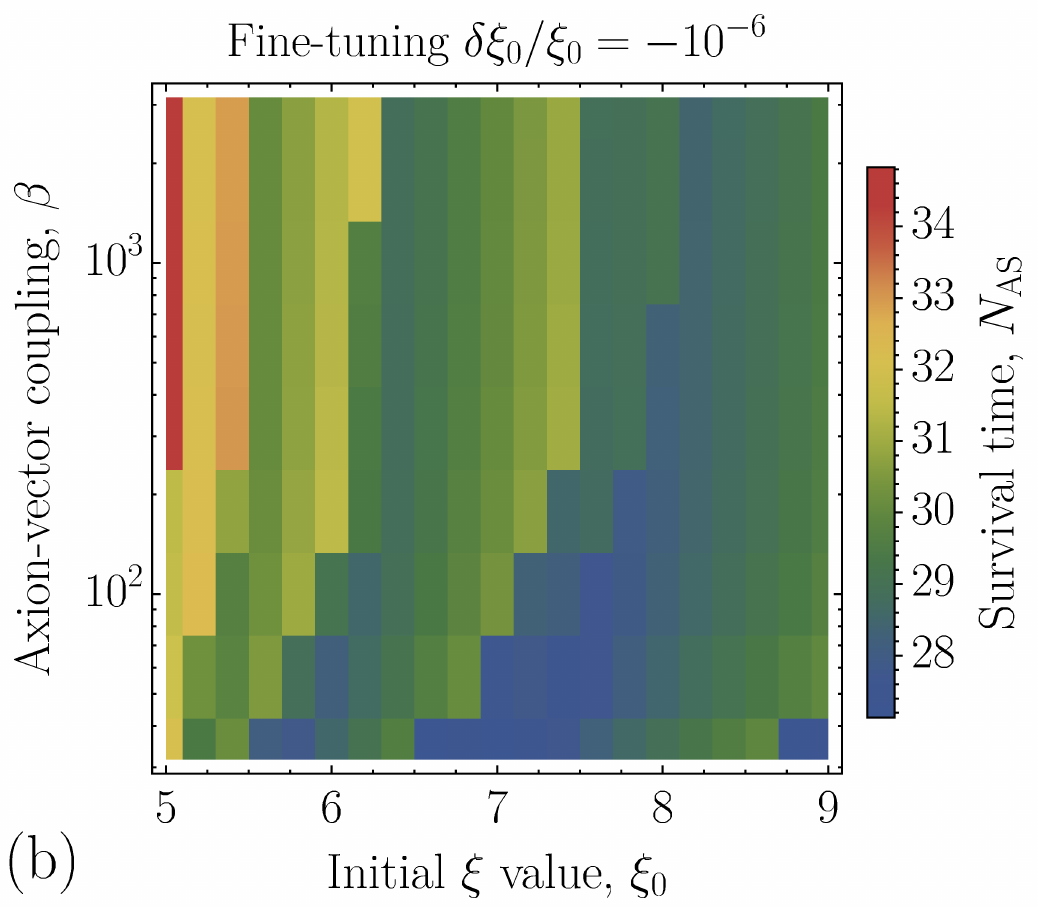}\\
\vspace{3mm}
\includegraphics[width=0.48\linewidth]{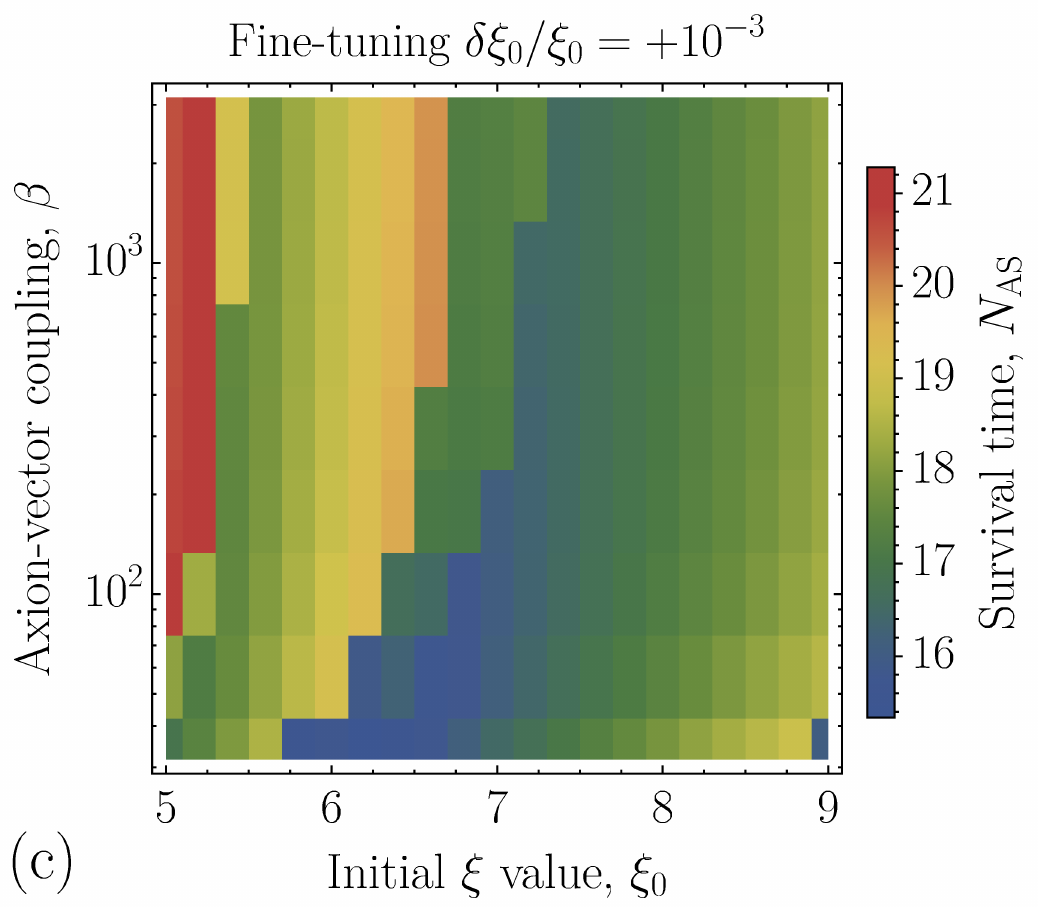}
\hspace{3mm}
\includegraphics[width=0.48\linewidth]{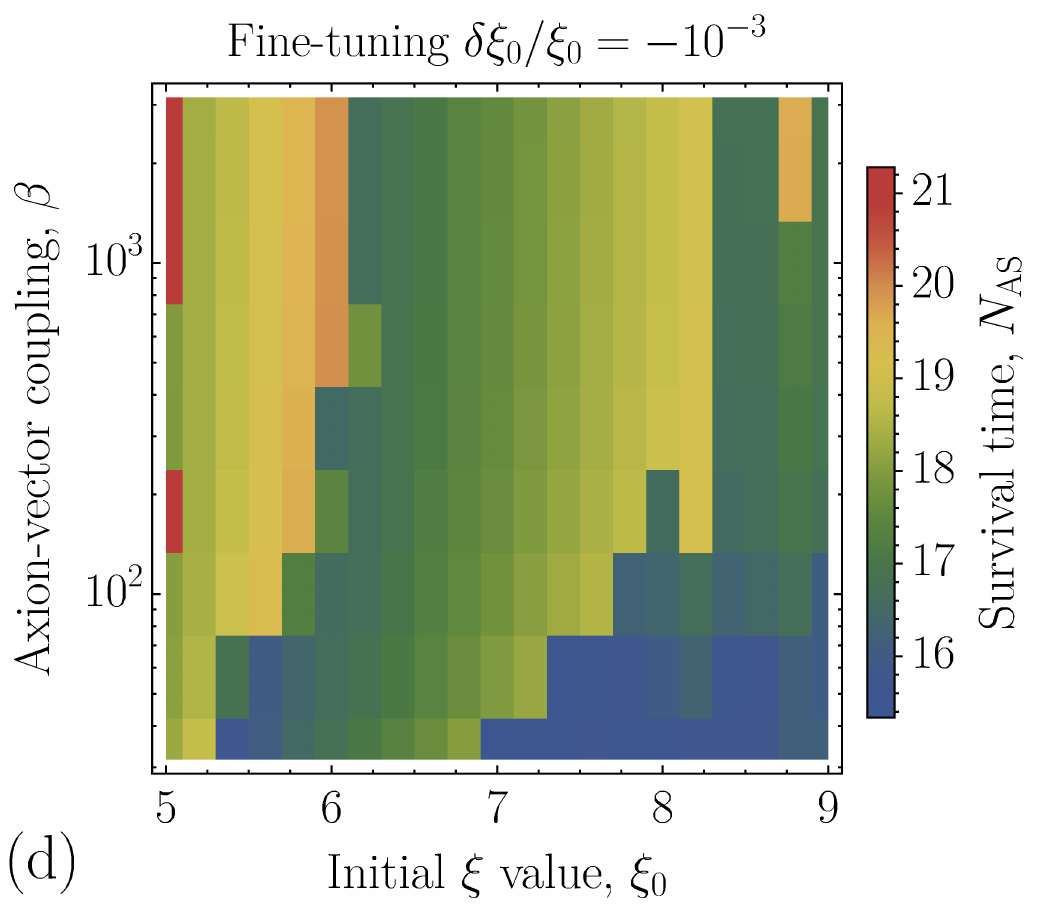}
\caption{Survival time of the AS solution, $N_{\rm AS}$, as a function of the parameters $\xi_0$ and $\beta$ for four values of $\xi$ fine-tuning: (a) $\delta\sm\xi_0/\xi_0=+10^{-6}$, (b) $\delta\sm\xi_0/\xi_0=-10^{-6}$, (c) $\delta\sm\xi_0/\xi_0=+10^{-3}$, (d) $\delta\sm\xi_0/\xi_0=-10^{-3}$.}
\label{fig-heatmap}
\end{figure}

In order to see how the survival time of the AS solution depends on the model parameters, the axion--vector coupling constant $\beta$ and the initial production parameter $\xi_0$, we perform a scan over this two-dimensional parameter space and present the results in the form of heatmap plots in Fig.~\ref{fig-heatmap}. We fix the initial relative deviation in $\xi$ to be (a)~$\delta\sm\xi_0/\xi_0=+10^{-6}$, (b)~$-10^{-6}$, (c)~$+10^{-3}$, and (d)~$-10^{-3}$,  which are shown in the corresponding panels of Fig.~\ref{fig-heatmap}. 
The comparison between the two top and the two bottom panels again confirms the scaling of Eq.~\eqref{eq:NASestimate-v2}. For ${\rm Re } \left( \zeta_1 \right) \simeq 0.55$, as indicated by Fig.~\ref{fig-zeta-on-xi}, a variation of $10^3$ in $\delta \sm\xi_0 / \xi_0$ provides a shift $N_{\mathrm{AS}} \simeq 12.5$, in good agreement with the various panels. More interestingly, each panel shows how the survival number of $e$-folds depends on the model parameters $\xi_0$ and $\beta$. This dependence is characterized by a wavelike pattern, meaning that the survival time changes non-monotonically with the increase of $\beta$ or $\xi_0$. This can be explained by the fact that the time dependence of $\delta\sm\xi/\xi_0$ is an oscillatory function with increasing amplitude [see Fig.~\ref{fig-evolution}(a)]. Typically, a slight change in the model parameters leads to a small phase shift in the oscillations, and the curve for $|\delta\sm\xi/\xi_0|$ crosses the threshold value of $10^{-1/2}$ at a slightly different time. There can, however, be situations where, after a small change in $\beta$ or $\xi_0$, the curve for $|\delta\sm\xi/\xi_0|$ does not reach the threshold during the same oscillation as before, but it has to evolve approximately half an oscillation period more to do this. This results in jumps in $N_{\rm AS}$, as can be clearly seen in Fig.~\ref{fig-heatmap}.

Overall, our results in Figs.~\ref{fig-survival-time} and \ref{fig-heatmap} corroborate our understanding of the relation between the growth rate of the fastest-growing mode, $\operatorname{Re}(\zeta_1)$, and the survival time of the AS solution, $N_{\rm AS}$, and thus serve as another numerical validation of the analysis in Sec.~\ref{sec:const_back}.


\section{Self-correction algorithm for the GEF}
\label{app:self-correction}

\begin{figure}[ht]
\centering
\includegraphics[width=0.775\linewidth]{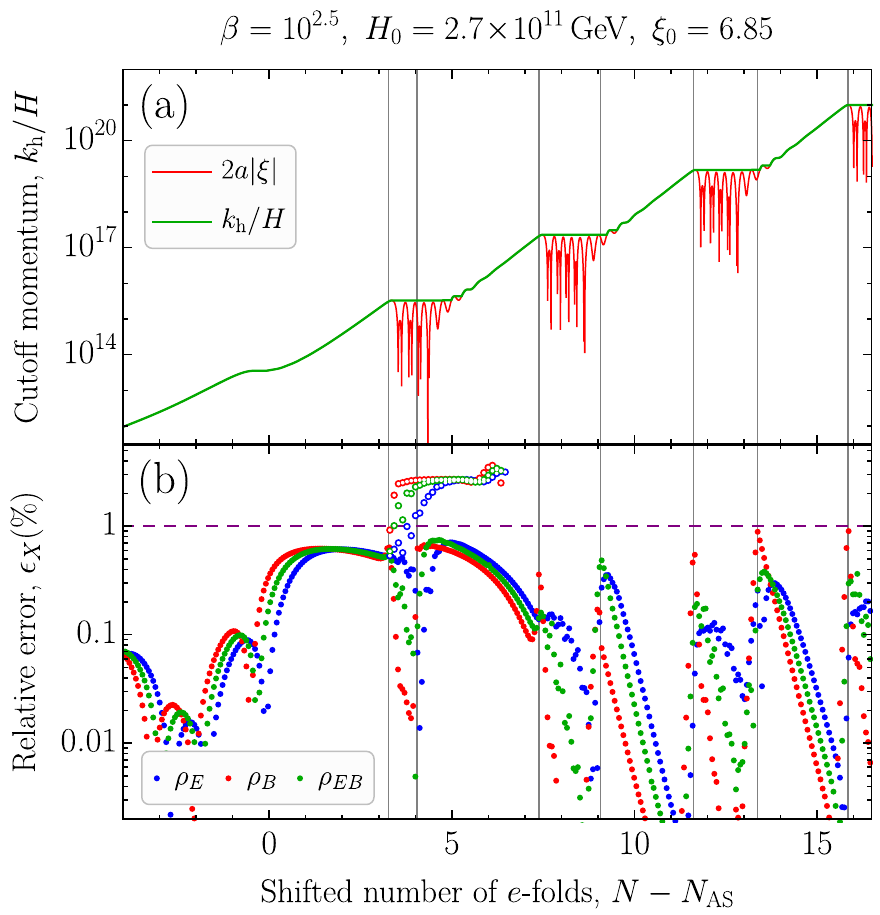}
\caption{Time dependences of (a)~the cutoff momentum $k_{\mathrm{h}}$ and (b)~the relative error of the GEF result for $\rho_E$, $\rho_B$, and $\rho_{EB}$ compared to the mode-by-mode (MBM) solution for $\beta=10^{2.5}$, $\xi_0=6.85$, and $H=2.7\!\times\! 10^{11}\,$GeV. Gray vertical lines show the moments of time when the self-correction procedure has been applied. The empty circles show the relative error of the GEF result compared to the MBM solution that one finds when no self-correction procedure is applied.}
\label{fig-selfcor}
\end{figure}

At the end of Sec.~\ref{sec:const_back}, we discuss a challenge that we encounter when employing the GEF during the period of fast oscillations shown, \textit{e.g.}, in Figs.~\ref{fig-late-time}, \ref{fig-SR-7}, \ref{fig-SR-6-85}, and \ref{fig-SR-6-85-compare}. For such a fast and non-monotonic change in $\xi$ (or, equivalently, in the inflaton velocity), the cutoff momentum given by Eqs.~\eqref{k-h-final} and \eqref{k-h-final-N} shows a sequence of plateaus in the time evolution. This is clearly seen in Fig.~\ref{fig-selfcor}(a), plotted for the case $\beta=10^{2.5}$ and constant background quantities $H_0=2.7\!\times\!10^{11}\,$GeV and $\xi_0=6.85$, where the red curve shows the expression $2\,e^{N}|\xi(N)|$, and the green solid line is the upper envelope of this function, which is $k_{\mathrm{h}}/H$.

During these plateau stages, the underlying assumption that the spectral densities of $\mathcal{E}^{(n)}$, $\mathcal{G}^{(n)}$, and $\mathcal{B}^{(n)}$ are dominated by the mode $k_{\mathrm{h}}$ at large $n$, which allows us to truncate the GEF system at some finite order $n_{\mathrm{cut}}$, is violated. Indeed, let us consider the mode equation \eqref{A_2}. In the case of constant $H$, it implies that, at a given moment of time $N$, the tachyonic instability occurs for modes with momenta
\begin{equation}
k <  2He^{N}\left|\xi(N)\right| \,.
\end{equation}
During each of the plateau stages, all these momenta are less than $k_{\mathrm{h}}$, which equals the largest value of $2He^{N}|\xi(N)|$ in all preceding moments of time. This means that the spectral densities are growing for modes with smaller momenta than $k_{\mathrm{h}}$, such that these modes may become of the same importance as $k_{\mathrm{h}}$ in the integrals over the spectra. This introduces a numerical error in the last equations of the GEF, where the truncation is performed, which then quickly propagates through the system of equations, finally reaching the zeroth order. Note that increasing $n_\mathrm{cut}$ does not help to avoid this problem but only postpones it to higher-order bilinear quantities. Therefore, in order to detect the situation where the GEF starts giving inaccurate results, one should always perform a consistency check using the mode-by-mode (MBM) solution. For this, one takes the time dependence of the scale factor $a$ and the inflaton velocity $\dot{\phi}$ from the GEF result [in the case of constant background quantities, one just needs to take the dependence $\xi(N)$], and solves the mode equation \eqref{A_1} or \eqref{A-const-xi-H} for all modes that cross the horizon during the time interval used in the GEF. Then, using Eqs.~\eqref{En-sp}--\eqref{Bn-sp}, one can compute the bilinear functions that follow from the MBM solution and compare them to those from the GEF result. The relative deviations between them,
\begin{equation}
    \epsilon_X \equiv \Big|\frac{X^{\mathrm{GEF}}-X^{\mathrm{MBM}}}{X^{\mathrm{MBM}}}\Big|\,, \label{error}
\end{equation}
where in the place of $X$ one may take, \textit{e.g.}, the lowest-order bilinear gauge-field quantities $\mathcal{E}^{(0)}$, $\mathcal{G}^{(0)}$, and $\mathcal{B}^{(0)}$, can be used to estimate the consistency of the GEF solution. Note that this relative deviation is not a true numerical error of the GEF result, because the MBM solution that we use as a reference is not independent from the GEF solution, but it is based on the time dependence of $\xi$ taken from the GEF result. We will, nevertheless, refer to it as an ``error'' in what follows.

The empty circles in Fig.~\ref{fig-selfcor}(b) show the typical behavior of the relative error during the plateau in $k_{\mathrm{h}}$: at some point, it starts increasing and exceeds the 1\,\% threshold (shown by the purple dashed line). If one allows the GEF system to go further in time, the error reaches much greater values. Therefore, in order to control the accuracy of the GEF results, one needs to reinitialize it at the moment when the error exceeds the selected threshold. To do so, one may use the spectra obtained by the MBM approach in order to compute the bilinear functions according to Eqs.~\eqref{En-sp}--\eqref{Bn-sp}. This helps to improve the situation and keep the error under control. The vertical gray lines in Fig.~\ref{fig-selfcor} mark the times at which the self-correction was performed. The error in the corrected result remains always less than the chosen threshold of 1\,\% during the whole duration of the simulation.

The algorithm underlying our self-correction procedure can be summarized as follows:
\begin{itemize}
    \item[(i)] Perform numerical runs of the GEF equations for two different values of $n_{\mathrm{cut}}$ (sufficiently large so that, for the time intervals without plateaus in $k_{\mathrm{h}}$, the results of both runs coincide)\footnote{In the absence of plateau regions, the error of the GEF result first decreases with increasing $n_\mathrm{cut}$ and then tends to a certain constant residual value which is caused not by the truncation procedure but by other approximations, \textit{e.g.}, in the boundary terms, see Ref.~\cite{Gorbar:2021rlt}. Therefore, it is important to compare the results with two different values of $n_{\mathrm{cut}}$ and make sure that the error introduced by the truncation procedure is negligible.} up to the time in which the results start to deviate.
    \item[(ii)] Use the time dependence of $\xi$ and $a$ from the GEF result and solve the mode equation \eqref{A_1} in order to obtain the mode spectrum of the produced gauge fields.
    \item[(iii)] Compute the zeroth-order bilinear quantities from the spectrum using the expressions in Eqs.~\eqref{En-sp}--\eqref{Bn-sp} and find the relative error of the GEF result using Eq.~\eqref{error}.
    \item[(iv)] When the error exceeds the set threshold, compute the values of the bilinear functions according to Eqs.~\eqref{En-sp}--\eqref{Bn-sp} for $n>0$.
    \item[(v)] Use these new corrected values for the bilinear functions to reinitialize the GEF.
\end{itemize}

In order to avoid small jumps (by $\sim 1\,\%$) in the zeroth-order quantities, $\mathcal{E}^{(0)}$, $\mathcal{G}^{(0)}$, and $\mathcal{B}^{(0)}$, which may lead to a short spurious stage of relaxation to a smooth solution just after the reinitialization, it is better to reinitialize only the bilinear quantities starting from $n=1$, while keeping the old values for the zeroth-order quantities along with the values of $\xi$ and $a$, which cannot be updated by the MBM approach. Finally, we comment on the choice of the threshold in the error. If one chooses a greater (smaller) threshold, less (more) frequent self-corrections are required. In practice, one therefore needs to find a compromise between the accuracy of the numerical result and the required computation time.

\vspace{5mm}
\noindent\textbf{Open Access.} This article is distributed under the terms of the Creative Commons Attribution License (\href{https://creativecommons.org/licenses/by/4.0/}{CC-BY 4.0}), which permits any use, distribution and reproduction in any medium, provided the original author(s) and source are credited.


\bibliographystyle{JHEP}


\end{document}